\documentclass[ twocolumn,
aps,prd,   
               preprintnumbers,numbers,sort&compress,nofootinbib,
                            showpacs,
               colorlinks,
               linkcolor=blue,   
               citecolor=blue]{revtex4-1}
   \newcommand{\exclude}[1]{}

\usepackage{graphicx,amsmath,amssymb,bm}
\usepackage{psfrag}
\usepackage{feynmp}
\usepackage{hyperref}
\usepackage{enumitem}

\newcommand{\beq}{\begin{equation}}
\newcommand{\eeq}{\end{equation}}
\newcommand{\be}{\begin{eqnarray}}
\newcommand{\ee}{\end{eqnarray}}

\def\dd{ \,\mathrm{d} }

\def\+{\dagger}
\def\la{\langle}
\def\ra{\rangle}
\def\<{\langle}
\def\>{\rangle}

\begin{document}

\title { Superfluid  helium II  as the QCD vacuum }
 
 \author{Ariel Zhitnitsky}
 
  \affiliation{~Department of Physics \& Astronomy, University of British Columbia, Vancouver, B.C. V6T 1Z1, Canada}


\begin{abstract}
We study 
the winding number susceptibility  $\la I^2\ra$ in a superfluid system and the topological susceptibility  $\la Q^2\ra$   in QCD. 
We argue that both correlation functions exhibit similar   structures, including the generation of the contact terms. We discuss the nature of the contact term in superfluid system  and argue that  it has exactly the same origin as in QCD, and it is related to  the long distance physics which cannot be associated with conventional microscopical degrees of freedom such as phonons and rotons.    We emphasize  that  the conceptual similarities  between superfluid system and QCD may lead, hopefully,   to a deeper understanding of the topological features of a superfluid system as well as the QCD vacuum. 
\end{abstract}

\maketitle

\section{Introduction. Motivation.}
The  main goal of this work is to present few arguments suggesting that  a superfluid system   has a number of features which are normally attributed to the QCD vacuum.  In other words, while QCD is a system with a  gap, it still exhibits some phenomena  which are typically present in   the systems with long range order such as superfluid liquid.  

The basic objects of our study are the winding number susceptibility  $\la I^2\ra$ in superfluid system and the topological susceptibility  $\la Q^2\ra$   in QCD. The reason for our focus on these correlation functions is that  the superfluid density $n_s$  can be    expressed in terms of the correlation function $\la I^2 \ra$, while the  vacuum energy in QCD is explicitly expressed in terms of $\la Q^2\ra$. Furthermore,   the topological susceptibility  $\la Q^2\ra$ plays the crucial role in resolution of the celebrated   U(1) problem in QCD. 
  Computation of such  type of correlation functions   is very hard technical problem which includes the dynamics of the non-local  winding number operator as well as the dynamics of microscopical degrees of freedom   carrying  the vorticity.  Fortunately,   the topological susceptibility  in QCD   has been extensively studied in strongly coupled QCD  and other gauge field theories with nontrivial topological features. The experience from these QCD studies  may give us a hint about the behaviour of  the winding number susceptibility  $\la I^2\ra$ in superfluid systems.  Such an analogy   (applying     in the opposite way, from a superfluid system   to QCD)  may provide us with some new ideas   on the  nature of the phase transition in strongly coupled gauge theories when $\la Q^2\ra$ experience some drastic changes according to the lattice studies.

  We argue that  these correlation functions demonstrate  very  similar features. In particular, they both exhibit the  contact terms which are  originated from the long distance dynamics not associated with any microscopical local propagating degrees of freedom such as well-studied in superfluidity phonons or rotons. This contact term is known to play a key role in the resolution of the $U(1)$ problem in QCD. We elaborate on  properties of a similar contact term in superfluid systems.  
 
      The history of physics has a long list  of examples when the conceptual similarity between particle physics and condensed matter systems benefits both fields. In the present work we hope to extend  this long list by adding  one more example where  the topological susceptibility $ \la Q^2\ra$  in QCD and  correlation function $\la I^2 \ra$ in superfluid systems both exhibit  similar and very unexpected properties.

Our presentation is organized as follows. In next few sections \ref{winding0}, \ref{eff_lagrangian} we introduce our notations and definitions related to the topological properties of a superfluid liquid. We also argue that conventional Landau criterion  cannot be used as a criterion for superfluidity. Rather a different  criterion for superfluidity  has to be used, and it  should be  formulated   in terms of the winding number, 
rather than in  terms of phonon- roton dispersion relation.  The corresponding arguments will be reviewed in    section  
\ref{criterion}.   

In  section \ref{vortices} we introduce auxiliary gauge field to describe the vortices and circulation in superfluid systems, similar to vector gauge potential in $E\&M$ theory.
 In sections  \ref{susceptibility}, \ref{numerics}  we express the partition function and the winding number susceptibility $\la I^2\ra$   in terms of these auxiliary topological gauge fields.   Finally, in section \ref{interpretation} we elaborate on a number of similarities and differences between our computations  of $\la I^2\ra$ for superfluid liquid and the computations of $\la Q^2\ra$  in QCD. 
 \exclude{While  there is a formal similarities in structure between these two correlation functions there is  a fundamental difference between  the two: a  complicated network of twisted, knotted and wrinkled superfluid vortices  is  defined in Minkowski space time, while
 complicated topological configurations saturating  $\la Q^2\ra$ are defined in  Euclidean 4d space-time }
 We formulate the main lessons of our analysis  in concluding section  \ref{conclusion}.    We also speculate there on possible relevance of our studies for understanding the nature of the observed cosmological  dark energy, as the vacuum energy in the system is directly related to  the contact term in   $\la Q^2\ra$ in QCD  (and $\la I^2\ra$  in a superfluid system)  which is the main object of our  studies  in the present   work.

\section{\label{winding0}Winding number   and its properties}

 In what follows we specifically discuss a bosonic liquid such as $^4He$ to avoid any additional complications related to the fermionic structure of $^3He$ and its additional topological structures. One should comment that it is normally assumed the superfluid velocity is curl-free in superfluid $^4He$, i.e. $\vec{\nabla}\times \vec{v}_s=0$. 
 However, on a large scale the motion of superfluid $^4He$ is not really irrotational even when the density of the normal component is very small. In fact, the corresponding circulation is quantized, 
 \be
 \label{circulation}
 \oint \vec{v}_s\cdot d\vec{r}= \int d\vec{S}\cdot (\vec{\nabla}\times \vec{v}_s)=\frac{2\pi l\hbar}{m}= l\kappa_0,  
  \ee
  where $\kappa_0\equiv \frac{2\pi \hbar}{m}$ is unit flux of circulation and $l$ is integer. 
  It is important to emphasize that the quasiparticles, the phones and rotons do not transfer energy directly to and from the superfluid component. Nevertheless, the interaction between the two components can be observed indirectly as the quasiparticles can scatter off the vortex lines. Eventual  manifestation of this scattering is mutual friction between the components.  
  
 The key observation for our present discussions can be explained  as  follows. We want to express the superfluid density $n_S$ in terms of specific correlation function  $\la I^2\ra$ formulated in terms of the  winding number    ${I}$ as explained  below section \ref{criterion}. The relevant for our discussions object  is  defined as follows \cite{Pollock:1987zz,Ceperley:1995zz, book},
  \be
 \label{I-definition}
 I=\oint_{\Gamma} \vec{\nabla}\alpha\cdot d\vec{l}=2\pi l, ~~ \vec{I}\equiv \int_{\mathbb{M}}  d^3x \vec{\nabla}\alpha.
 \ee
   In formula (\ref{I-definition})  the function   $\alpha (x)$ describes the Nambu-Goldstone degree of freedom which itself is   the  phase of a scalar complex field $\Phi=\sqrt{n}\exp (i\alpha)$. A manifold $\mathbb{M}$ in definition  (\ref{I-definition})  is assumed 
  to have at least one  non-contractible path $\Gamma$ such that there is at least one nontrivial mapping $\pi_1[U(1)]=\mathbb{Z}$ between  $U(1)$ phase $\alpha$  and path $\Gamma$.  In case of  a 3-torus   $ \mathbb{T}^3$  there are  3 different slices describing 3 different mappings  $\pi_1[U(1)]=\mathbb{Z}$ for each slice, such that the system is characterized by 3 different components of vector $\vec{I}$.

  The  definition (\ref{I-definition})   is   very similar in structure to the conventional topological winding number  in two-dimensional gauge field 
 theories,   such as the 2d Schwinger model,   with nontrivial mapping $\pi_1[U(1)]=\mathbb{Z}$.   
  It is assumed that the periodic boundary conditions are imposed on the angular variable $\alpha$ in definition (\ref{I-definition}), and therefore the topological invariant 
  $\vec{I}$ is conserved.    
   
 Few comments are in order. 
 First, for the winding number $\vec{I}$ to   change,  an entire path across the periodic cell must   change.  In non relativistic superfluid it could only happen as a result of tunnelling transitions 
  (which is negligibly small effect  for macroscopically large systems) or interactions of the system with fluctuating  vortices. The reason why vortices may change the winding number is that  they  
 are characterized (locally) by vanishing superfluid density $n_s\simeq 0$ inside  the vortex core, where the winding number can  locally ``unwind" itself. Such an interaction can in principle change  the   topological invariant (\ref{I-definition}) by transferring it to the   vortices. Therefore, our comment is that the coupling of the winding state classified by $\vec{I}\neq 0$  with the vortices may, in principle,    transfer the winding number from the bulk to  a boundary.  It could only happen when a sufficiently large number of coherent   vortices represented by a  macroscopically large proliferated  vortex loop  is  present in the system \cite{book}. It is quite obvious that  at  zero, or very low  temperature, when there are very few   vortices  present in the system the topological invariant (\ref{I-definition}) is conserved  as the superfluidity is protected by the topological arguments.  
 Only at sufficiently large temperature at $T\simeq T_c$ a  percolated vortex network may emerge and remove the winding number to the boundary. It is exactly
 the temperature when the phase transition occurs. 

  This picture when the winding number $\vec{I}$ is conserved should be  contrasted with relativistic quantum field theories, such as QCD,  where the  tunnelling transitions constantly and continuously occur all the time, selecting a specific $|\theta\ra$ vacuum state, in contrast with non relativistic systems where the ground state is normally classified by the conserved winding number $|l\ra$, rather than by $\theta$ parameter representing a superposition of different winding states.

  In fact, as emphasized in \cite{book} the phenomenon of superfluidity itself is almost a trivial consequence of topological features of a complex scalar field defined on a nontrivial manifold  $\mathbb{M}$ with at least one  nontrivial mapping $\pi_1[U(1)]=\mathbb{Z}$. This   scalar field satisfies the Gross-Pitaevskii equations and effectively describes a superfluid liquid, as will be reviewed  in next section \ref{eff_lagrangian}.   
 
To reiterate: the superfluidity itself is a relatively simple   problem of  {\it classical} field theory of  a scalar complex field governed by Gross-Pitaevskii  Lagrangian. 
 A hard problem in this system is understanding of   the mechanisms of  how the superfluidity is getting destroyed 
  by some quantum or thermal fluctuations.  
  The relevant dynamics must include, in one way or another,  the    fluctuations which carry the topological vorticity and which are capable to remove or destroy the winding number,  as explained above.  
  Precisely this problem on possible mechanisms of the phase transition  between a normal $\vec{I}=0$ and superfluid $\vec{I}\neq 0$ states  is the main subject of the present work.

 \section{\label{eff_lagrangian}  Superfluidity: Phonons and Rotons. }
 We now want to review    the   well known portion  of the effective Lagrangian describing the   Goldstone modes related to spontaneous violation of the global $U(1)$ symmetry. The starting point  is Gross-Pitaevskii description when the superfluidity  is described in terms of  a single  scalar complex field $\Phi$ with non-vanishing  expectation value $\la\Phi\ra=n_S$, while its phase describes the corresponding Goldstone boson 
 \be
 \label{GP}
 L_{GP}&=&-\frac{i}{2}\left(\Phi^*\partial_t \Phi-\Phi\partial_t\Phi^*\right)\nonumber\\
&-& \frac{1}{2m}|\vec{\nabla}\Phi |^2 
 -\frac{\lambda}{2}\left(|\Phi|^2-n_S\right)^2. 
 \ee
 We define the Goldstone boson as the phase of the scalar field $\Phi=\sqrt{n}\exp(i\alpha)$ such that the relevant part of the Lagrangian assumes the form
 \be
 \label{GP1}
 L_{GP}= \left(n\partial_t \alpha \right)-  \frac{n}{2m}(\vec{\nabla}\alpha)^2 
 -\frac{\lambda}{2}\left(n-n_S\right)^2. 
 \ee
 Integrating out the heavy $n(t, \vec{r})$ field, which corresponds to substitution 
 \be
 \label{GP2}
 n\simeq n_S+\frac{1}{\lambda} \left[\partial_t \alpha -  \frac{1}{2m}(\vec{\nabla}\alpha)^2\right], 
 \ee
  leads to the following conventional Lagrangian describing the massless Goldstone field $\alpha (t, \vec{r})$
 \be
 \label{G}
 L_{G}= \left(n_S\partial_t \alpha \right) +\frac{1}{2\lambda} (\partial_t \alpha)^2-  \frac{n_S}{2m}(\vec{\nabla}\alpha)^2 +{\rm interactions.}~~~~~ 
 \ee
The quadratic   terms in eq. (\ref{G}) describe  the massless Goldstone field $\alpha (t, \vec{r})$
 with dispersion relation $\omega\sim k$.  
 The velocity field $\vec{v}(t, \vec{r})$ introduced earlier is related to the Goldstone field as follows, $\vec{g} (t, \vec{r}) =m\vec{v}=\vec{\nabla}\alpha$. As a result of this relation one can check  that the behaviour of the winding number $I_i$  can be expressed in terms of the velocity field
 as equation (\ref{I-definition}) states.
 
  A generalization of this  analysis  when the entire background is  slowly moving with nonzero velocity $\vec{v}_{\rm slow}\neq 0$ or experience a rotation with $\Omega\neq 0$ is straightforward. The resulting   effective Lagrangian can be written in terms  for the massless Goldstone fields propagating in  the curved background  
 \be
 \label{G1}
 L_{G}=  \frac{1}{2}\sqrt{-g} g^{\mu\nu}\partial_{\mu}\alpha\partial_{\nu}\alpha +{\rm interactions}, 
 \ee
   where $ g^{\mu\nu}$ is the so-called induced acoustic metric which can be explicitly computed in terms of the original parameters of the theory \cite{Volovik:2000ua}.
    
 The first term in (\ref{G}) is a total derivative and does not change the equation of motion for the Goldstone $\alpha  (t, \vec{r})$  field. However, this term can not be ignored as it will rise to a topological phase.  In fact, in all respects this term is similar to the Berry phase.   Furthermore, one can show that this term can be interpreted as a source for the Magnus forces for a moving vortex \cite{Volovik:2000ua,Stone}. Intuitively, this  is    expected result as   first term in  (\ref{G})  is similar, in all respects,  to  the phase factor that would be generated by a charged particle moving in a uniform magnetic field with the action $e\oint  dt\dot{x}_{i}A_{i}$. 
 
 The   phonons reviewed above are  not the only   quasi-particles  present in the  system. 
 Another type of quasi-particles, are   the so-called rotons \cite{Landau,Feynman}. The roton's    properties can be studied by analyzing the so-called form-factor
   $S(\vec{k})$  defined as follows
 \be
 \label{S}
 S(\vec{k}) =\int d^3 r \delta n (\vec{r}) e^{i\vec{k}\cdot\vec{r}}, ~~~~\delta n \simeq \left(n-\la n\ra\right)
 \ee
 where $ \delta n (\vec{r}) $ describes  the density fluctuations  of atoms in the liquid,  which in low energy effective description  can be approximated  by 
 the effective field $\Phi$ according to eq. (\ref{GP}). In terms of form factor $S(\vec{k})$ the spectral properties of excitations can be expressed as follows 
 \be
 \label{epsilon}
 \epsilon (\vec{k})=\frac{\hbar^2 \vec{k}^2}{2m S(\vec{k})}.
 \ee
  The form factor $S(\vec{k})$ is known experimentally from neutron scattering. It has the following features.  For sufficiently small $|\vec{k}|$ the form factor $S(\vec{k})$ shows a linear scaling, $S(\vec{k})\sim |\vec{k}|$. It  can be identified with excitations of phonons (\ref{G}), see  original papers \cite{Landau, Feynman} and the textbook \cite{book} with nice historical comments. For linear $S(\vec{k})\sim |\vec{k}|$ the    dispersion relation (\ref{epsilon}) indeed exhibits  the linear scaling $\epsilon (\vec{k})\sim  |\vec{k}| $  consistent with phonon's interpretation.     Another profound feature of the form factor $S(\vec{k})$ is that it has a maximum at wave number $  |\vec{k}_0| \simeq 2 ~ \mathring{A}^{-1}$. In vicinity of this maximum one can expand $S(\vec{k}) \simeq S_0(k_0)-\frac{1}{2}|S_0^{''}|(\vec{k}-\vec{k}_0)^2$ such that $ \epsilon (\vec{k})$ exhibits    the behaviour corresponding to the gapped modes with dispersion relation
  \be
 \label{epsilon1}
 \epsilon (\vec{k})\simeq \epsilon_0+\frac{\hbar^2 \vec{k}_0^2 |S_0^{''}|(\vec{k}-\vec{k}_0)^2}{4m S_0^2}, ~~~  \epsilon_0\simeq \frac{\hbar^2 \vec{k}_0^2}{2m S_0}.
 \ee
  The corresponding gapped excitations  have been  identified with rotons \cite{Landau, Feynman,  book}. It is important to emphasize that the gap $\epsilon_0$
  related to the rotons is a temperature dependent parameter, i.e.  $\epsilon_0(T)$. However,  while $\epsilon_0(T)$ slowly varies with the temperature,  
  it does not vanish at the critical temperature $\epsilon_0(T_c)\neq 0$ and remains approximately constant $\epsilon_0(T\gg T_c)\approx 5K$
  for the temperatures well above the critical $T_c$, see e.g.\cite{Bobrov} for references on numerous experimental results. 
  This property will play an important role in our following discussions, where we argue that another dynamical gap parameter emerges in the system, which however vanishes at $T=T_c$, and therefore cannot be identified with  conventional roton's gap $\epsilon_0(T)$. 
 
 The rotons play a key role in formulation of the
 Landau criterion 
 \be
 \label{Landau}
 \omega_k+\vec{v}\cdot \vec{k} >0, ~~  ~~( \text{Landau~ criterion})
 \ee
 which is commonly interpreted in the literature as a criterion for superfluidity. Furthermore, there is a wide spread opinion that the rotons characterized by typical dispersion relation (\ref{epsilon1}) exist only in superfluid states. 
 
 There is a number of arguments why this interpretation cannot be correct. First of all, there is a numerous  experiments which convincingly show that the phonon-roton spectrum   is not unique for superfluidity, but in fact, is very generic characteristic of a liquid state. 
 Examples include, but not limited to such systems as  liquid titanium, normal (not superfluid) helium, molecular para-hydrogen, neon, oxygen in supercritical region, and many others, see e.g. \cite{Bobrov} for references on the experimental results. Furthermore, it is well-known fact that the critical velocity  calculations based on the measured roton's minimum is much higher than observed values by orders of magnitude. 
 
  An independent  argument \cite{book} which basically leads  to the same conclusion  is based on topological reasoning suggesting that the topological invariant (\ref{I-definition}) remains intact   as long as large vortex loop is not generated in the system. The Landau criterion (\ref{Landau}) obviously does not  carry  any information about the large vortex loops in the system as it is formulated in terms of the local microscopical degrees of freedom:   the rotons and phonons. Therefore, the  topological arguments  \cite{book} also suggest that  eq.(\ref{Landau}) cannot serve 
 as a criterion for superfluidity. 
 
 \section{novel criterion for superfluidity}\label{criterion}
 In our studies in the present work we shall use a different criterion for the superfluidity based on the correlation function for the winding numbers $\la I^2\ra$ defined in terms of the  topological invariant (\ref{I-definition})  with non-contractible  $\Gamma$. The key observation  of refs. \cite{Pollock:1987zz,Ceperley:1995zz, book} is that the corresponding correlation function is directly related  to the superfluid density $n_S$, and therefore, can serve as a criterion for superfluidity. In what follows in all our constructions we always assume that we are dealing with non-contractible  $\Gamma$ when the topological invariant (\ref{I-definition})  is conserved.  
 
 The conservation of $\vec{I}$ implies that one can introduce an intensive variable $\vec{s}$ which is thermodynamically conjugated  to $\vec{I}$
 such that the grand canonical potential can be represented as follows \cite{book}:
 \be
 \label{Z}
 \Omega( T, \mu, \vec{v}_n, \vec{s})=-T\ln {\cal{Z}}, ~~~ \la \vec{I}\ra=-\frac{\partial \Omega}{\partial \vec{s}}. 
 \ee
 The physical meaning of $\vec{s}$ can be inferred from the following generic relation  \cite{book}:
 \be
 \label{s}
 \vec{s}=n_s(\vec{v}_s-\vec{v}_n).
 \ee
 From this relation it follows  \cite{book} that in the reference frame of the walls where $\vec{v}_n=0$ one can express the superfluid density $n_s$ as the static response of the system with respect to small variation of $\vec{s}$:
 \be
 \label{n_s}
 \frac{1}{n_s}=-\frac{1}{3mV}\nabla_{s_i}^2\Omega  ~~~  {\vec{s}\rightarrow 0},
 \ee
 where the second derivative $\nabla_{s_i}^2\Omega$ can be explicitly 
 expressed as 
 \be
 \label{n_s1}
 \nabla_{s_i}^2\Omega= -\frac{1}{T} \la \vec{I}, ~ \vec{I}\ra, ~~~  {\vec{s}\rightarrow 0}
 \ee
  where the correlation function $\la \vec{I}, ~ \vec{I}\ra$ is defined as 
  \be
  \label{I2}
  &&\la \vec{I}, ~ \vec{I}\ra =   \lim_{\mathbf{k}\rightarrow 0}\int_{\mathbb{M}}  {d^3x} \int_{\mathbb{M}}e^{i\mathbf{k}\cdot(\mathbf{x}-\mathbf{x}')}    {d^3x'}  \\ 
 &&\cdot\la  {\vec{\nabla}} \alpha(\mathbf{x})  ,  {\vec{\nabla}}\alpha(\mathbf{x}') \ra .  \nonumber 
  \ee
  where factor $ \lim_{k\rightarrow 0}$ must be included into the definition to keep only connected portion of the correlation function, see few comments at the end of this section. As we discuss below, a proper evaluation of the contact term of the correlation function (\ref{I2}) at $\Delta \mathbf{x}\equiv (\mathbf{x}-\mathbf{x}')\rightarrow 0$ 
   requires  that both $\mathbf{x}$ and $\mathbf{x}'$ lie on the same   path $\Gamma\in \mathbb{M}$  when $\Delta \mathbf{x} $ approaches zero.  
     
  The expectation value in this expression should  be computed using grand canonical potential defined by (\ref{Z}). 
    Similar relation is known to exist  in QCD where the vacuum energy $E_{\rm vac}$ is expressed in terms of the topological susceptibility $(\partial^2 E_{\rm vac}/\partial \theta^2)\sim \la Q^2\ra$ where $Q$ is the topological density operator, and parameter $\theta$ is conjugated to $Q$ enters the QCD partition function as $\theta Q$. This QCD relation is the direct analog of eq. (\ref{n_s}) when the vacuum energy $E_{\rm vac}$ in QCD plays the role of $\Omega$, the  topological susceptibility  $\la Q^2\ra$ is analogous to the winding correlation function (\ref{I2}), while the QCD parameter $\theta$ is analogous to intensive variable $\vec{s}$. We want to see if this analogy is sufficiently deep, and whether the correlation function $\la I^2\ra $ in superfluid state   has similar  features which are known to be present in  the topological susceptibility in strongly coupled QCD.

  In what follows we want  to be more specific and  consider a simple  manifold $\mathbb{M}=\mathbb{S}^1\times \mathbb{I}^2$ which represents a hollow    cylinder length $L_z$, internal radius  $R_1$ and external radius $R_2$. In other words, for simplicity we consider a manifold which has a single nontrivial mapping  $\pi_1[U(1)]=\mathbb{Z}$ corresponding to one type of non-contractible  contour $\Gamma$ along $\mathbb{S}^1$. For this topology
  the winding number of the system has a single nontrivial component and can be represented as follows
   \be
  \label{I3}
  {I_z} =L_z(R_2-R_1) \int_{\mathbb{S}^1}  \partial_i\alpha d{l}_i=2\pi n^{\rm class}_z L_z(R_2-R_1), ~~~~
  \ee
where $n^{\rm class}_z\in \mathbb{Z}$  has the physical meaning of strength of vorticity (number of times the phase field $\alpha(x)\in U(1)$ makes a full circle   after traversing a closed path  along $\mathbb{S}^1\in \mathbb{M}$).  Factor  $2\pi L_z(R_2-R_1)$ represents the surface area of a hollow cylinder and can be interpreted as a corresponding degeneracy factor. Finally, we use subscript $n^{\rm class}$ to  emphasize  that this integer number classifies exclusively the classical portion of the winding number. 

The winding (\ref{I3}) is a conserved quantity  as we already explained. However, for the purposes of the present work we want to study the  
fluctuations of the winding number $\delta \vec{I}$ rather than its classical conserved portion (\ref{I3}).  This is because the fluctuations of the vorticity may potentially
change the winding number (\ref{I3}), which is precisely the main goal of our studies. Exactly  these fluctuations of the vorticity in form of generation of a large vortex loop will eventually lead to the phase transition at $T_c$.  

Formal manifestation of the quantum nature of  the correlation function (\ref{I2}) which is the main focus  of our studies,  is expressed in terms of    a factor  $ \exp({ikx})$ under the integral. The corresponding   $ \lim_{k\rightarrow 0} $ 
  should be evaluated      at the end of the computations. This formal procedure  selects the connected portion of this correlation function  entering  expression (\ref{n_s1}). 
  
  We believe that few historical comments are warranted here. First of all, the fact that configurations which are responsible for the phase transition must carry the vorticity and must be macroscopically large has been noticed (on the intuitive level)   long ago   by Feynman \cite{Feynman}. A   precise definition of superfluidity  in terms of the response of the  energy with respect  to variation of the boundary conditions around a nontrivial cycle $\Gamma$ was given by Leggett \cite{Leggett}.   In fact,  the definition formulated in  \cite{Leggett} is essentially identically the same  as  (\ref{n_s}) with the only difference that previously 
 it was analyzed in terms of  the  $N$ body wave function  in contrast with the modern definition  (\ref{n_s1}), (\ref{I2})   expressed   in terms of the effective long range  field $\alpha (\mathbf{x})$.  Finally,  Pollock and Ceperly   \cite{Pollock:1987zz,Ceperley:1995zz}  reformulated this property of superfluidity in terms of the  path integral.
   In our present analysis  we adopt the  modern definition \cite{book} of the    superfluidity  (\ref{n_s}),  (\ref{n_s1}), (\ref{I2})  being expressed    in terms of the correlation function $\la \vec{I}, ~ \vec{I}\ra$  which itself is formulated   in terms of the path integral.
     
  Precisely such a formal path integral representation  allows us to observe a deep   analogy between superfluid systems and  QCD, which is the main subject of this work. 
  The only additional comment we would like to make here  is that  this   analogy   will be  formulated in terms of the ``dual" variables to be  introduced in   section \ref{susceptibility}. It is much harder to understand   the same similarities   (which manifests itself as the long range order) in these two different systems in terms of the original   physical variables, see also related comment at the very end of concluding section \ref{relation}.

  \section{The    topological  vortices}\label{vortices}
    As we explained in previous section the main subject   of our studies is the physics of   fluctuating vortices. We want to avoid confusion with notations in what follows and define a different winding number $W_i$ which is the operator describing the quantum/thermal  fluctuations of the vorticity, 
    in contrast with classical expression  (\ref{I3}) which represents a conserved topological number of the system. These two objects, of course, are related to each other with a  simple  numerical factor representing the size of the system. However, we prefer to use a different normalization for the vortices in our discussions  in this section  to avoid some numerical dimensional factors such as size of the system entering (\ref{I3}). 
    For thermally excited  vortices to be studied below this size is essentially a microscopical size of the vortex's core rather than macroscopical size of a system. The vortices become very large in size, of the order of the system,   when $T$ approaches the critical value $T_c$. 
    
  Our definition for the winding number $W_z$ let us say, in $z$ direction assumes that we measure the vorticity by computing the contour integral along path $\Gamma$ which lies on the $xy$ plane such that the computations are reduced to evaluation of the  surface  integral $\partial S=\Gamma$. In particular, for a single vortex pointing along $z$ direction it is given by
  \be
  \label{vorticity1}
  W_z &=&\int_{\Gamma}  \partial_i\alpha (\vec{x}) \frac{d{l}_i}{2\pi} 
  =\int_S \frac{dx dy}{2\pi} (\partial_1\partial_2 -\partial_2\partial_1)\alpha(\vec{x}) \nonumber\\
   &=& n_z  \int_S dxdy\delta^2_z(x_{\perp}).  
  \ee
  The generalization of this topological description  (\ref{vorticity1}) for arbitrary vortex shape   is straightforward: 
  the winding number operator $W_k$ can be expressed in terms of the circulation field $\gamma_k (\vec{x}) $
as follows, 
\be
\label{W_new}
W_k=\frac{1}{2\pi}\int d{S}_k\gamma_k (\vec{x}), 
\ee 
 where $\gamma_k (\vec{x})$ is defined as 
 \be
 \label{vorticity2}
 \gamma_k (\vec{x})=\epsilon_{ijk}\partial_ig_j(\vec{x})=\epsilon_{ijk}\partial_i\partial_j\alpha(\vec{x}), 
 \ee
 where $g_j(\vec{x})=\partial_j \alpha(\vec{x})=mv_j(\vec{x})$ is the velocity field introduced earlier. 
In particular, for an   ideal structureless singular vortex (\ref{vorticity1}) one has 
 \be
 \label{vorticity3}
   \gamma_k (\vec{x})=    \epsilon_{ijk}\partial_ig_j (\vec{x})= 2\pi \delta^2_k(x_{\perp}).
 \ee
 The winding  $W_k$  counts number of crossing points between the vortices and the surface $S_k$.  
  The physical meaning of $\gamma_k (\vec{x})$ is quite obvious-- it describes the density of circulation per unit area  along the vortex  of arbitrary geometrical shape pointing  locally in $k$ direction. This function obviously satisfies the  conservation law,   $\partial_k  \gamma_k (\vec{x})=0$ for closed vortex lines even for the singular ones. This is a formal expression of the property  that the winding number (\ref{vorticity1}), (\ref{W_new}) does not depend on position of the cut along the vortex. If a vortex line is open with the ends at points $\vec{x}_i$ and $\vec{x}_f$ than 
 one can easy to see that $\partial_k  \gamma_k (\vec{x})=2\pi\left[\delta^3(\vec{x}-\vec{x}_i) -\delta^3(\vec{x}-\vec{x}_f)\right]$.  
 For closed vortices one has $\vec{x}_i=\vec{x}_f$ and one returns to the  conservation law, $\partial_k  \gamma_k (\vec{x})=0$.
 
One should emphasize that our fluctuating  vortices are not  straight lines all pointing in the same  direction, which is a conventional picture for rotating fluid. Rather, the picture we have in mind is that  the fluctuating  vortices look like a complicated dynamical mixture of knotted, crumpled and wrinkled    fluctuated spaghetti, with the only difference is that these spaghetti  make closed, rather then opened  vortices,  and  they also have   specific    chiralities.    

When we cut our spaghetti-like system, let us say, in $z$ direction, we observe a large number of plus and minus fluxes sitting on our  $(xy)$- slice   where we made a cut. A similar  picture would emerge if we  cut in $x$ or $y$ directions. This is precisely the reason why eq. (\ref{I-definition})  exhibits   three different conserved winding numbers $I_i$, and corresponding  conjugated variable $\vec{s}$ which enters the definition (\ref{Z}) is a 3-vector, rather than a scalar. It should be  contrasted with  non-abelian gauge QCD when there is a single topological classification scheme based on $\pi_3[SU(N)]=\mathbb{Z}$ and there is a single conjugated parameter $\theta$ which plays the same role as intensive parameter  $\vec{s}$ plays in superfluid systems.   
 
 It is very instructive  to present an analogy with magnetism as it gives a good intuitive picture about superfluid vortices and their description in terms of our Goldstone  field $\alpha(\vec{x})$,  velocity field $g_j (\vec{x})=\partial_j\alpha(\vec{x})$ and circulation density field $\gamma_k (\vec{x})=\epsilon_{ijk}\partial_i g_j(\vec{x})$. To be more precise, our field $g_i (\vec{x})$ behaves in all respects as the magnetic field $B_i(\vec{x})$.
 Indeed the magnetic field satisfies the following equations
 \be
 \label{magnetic}
 \partial_iB_i=0, ~~~\epsilon_{ijk}\partial_i  B_j(\vec{x}) =\mu_0 J_k(\vec{x}), ~~ \partial_kJ_k(\vec{x})=0.~~
 \ee
Equation $ \partial_iB_i=0 $ is similar to $ \partial_ig_i=m\vec{\nabla}\cdot \vec{v}_s=0$ which represents the incompressibility of superflow. At the same time equation $\epsilon_{ijk}\partial_i  B_j(\vec{x}) =\mu_0 J_k$ is similar to (\ref{vorticity2}), (\ref{vorticity3}) where 
 the role of $\mu_0 J_k$  plays  $ \gamma_k (\vec{x})$.
 The conservation of the current in magnetism $\partial_kJ_k(\vec{x})=0$ is expressed in terms of the property $\partial_k  \gamma_k (\vec{x})=0$ in our spaghetti -like system. The summary of this analogy can be represented in one line as follows
 \be
 \label{analogy}
 \partial_ig_i =0, ~~  \epsilon_{ijk}\partial_i  g_j(\vec{x}) = \gamma_k(\vec{x}), ~~\partial_k  \gamma_k (\vec{x})=0, 
 \ee
 which is formally coincides with (\ref{magnetic}). 
  The last equation  represents the invariance of the  winding number of the system $W_k$ with respect to   position of the cut  of the spaghetti environment which defines  a specific slice perpendicular to the $k$- direction.   The circulation field $\gamma_k (\vec{x}) $  in spaghetti system plays the   role of  the current distribution $J_k(\vec{x})$ in a  knotted  and twisted  system of wires in the magnetism. 
 
 One can make one more step in this formal analogy and introduce new  vector potential (in fact the axial potential) $a_j(\vec{x}) $ such that
 \be
 \label{a-potential}
 \epsilon_{ijk}\partial_i  a_j(\vec{x}) \equiv g_k(\vec{x}),
 \ee
 similar to $\vec{B}=\vec{\nabla}\times \vec{A}$.
 The field $a_j(\vec{x})$  obviously is not uniquely defined as the corresponding ``gauge transformation" $a_j(\vec{x})\rightarrow a_j(\vec{x})-\partial_j \Lambda(\vec{x})$ leave the physical velocity field $g_k(\vec{x})$ unaltered. Using the ``Coulomb" gauge $\partial_k a_k=0$ one can compute  the velocity  field  $g_k(\vec{x})$ in terms of the classical distribution of circulation sources  $\gamma_k (\vec{x}) $ as it is normally done in magnetism. To be more precise, the potential  $a_j(\vec{x}) $ according to eqs. (\ref{vorticity2}), (\ref{a-potential}) satisfies the following Poisson equation    
 \be
 \label{poisson}
 \vec{\nabla}^2 \vec{a}=-\vec{\gamma},  
 \ee 
 which has the following  solution  in terms of known Green's function
  \be
 \label{solution}
 a_k (\vec{x})=\frac{1}{4\pi}\int  d^3x' \frac{\gamma_k(\vec{x}')}{|\vec{x}-\vec{x}'|}.
 \ee 
 In case of infinitely thin single ideal vortex loop $\gamma_L$ with   unit  quanta (\ref{circulation}) of circulation formula (\ref{solution}) assumes the form 
  \be
 \label{solution1}
a_k (\vec{x}) =\frac{1}{2}\oint_{\gamma_L} dx'_k\frac{1}{|\vec{x}-\vec{x}'|},
 \ee
 where we used relation (\ref{vorticity3}) for $\gamma_k (\vec{x})$. In formula (\ref{solution1})  the contour   integral is taken along the vortex line $\gamma_L$. 
 
 The vector (axial) potential (\ref{solution1}) is long-ranged, similar to conventional vector potential $\vec{A}$ in magnetism.  This long-ranged interaction leads to the vortex-vortex interaction  similar to Biot-Savart law for the magnetism. To be more precise, for continuous distribution of circulation $\gamma_k(\vec{x})$ the interaction energy between vortices can be represented in the following form
 \be
 \label{v-interaction}
 E_{\rm int}=\frac{\sigma}{2}\int\frac{d^3x d^3x' }{4\pi} \frac{\gamma_k(\vec{x})\gamma_k(\vec{x}')}{|\vec{x}-\vec{x}'|},
 \ee
 where $\sigma={n_s}/{m}$ for an ideal infinitely long and infinitely thin vortex. In what follows we want to   treat $\sigma$ as a free parameter which determines the strength
 of vortex -vortex interaction with a finite size of the core. It is convenient for this purpose to introduce a dimensionless parameter $\eta\sim1$
 as follows  $\sigma=\eta {n_s}/{m}$ such that all deviations from the ideal case is coded by parameter $\eta$. It obviously depends on an internal vortex structure as well as on  its relation to a typical inter-vortex distance at temperature $T$.

 Using expression (\ref{solution}) for the vector potential $a_k (\vec{x})$  the formula for $E_{\rm int}$ can be also written in the local form
   \be
 \label{a-interaction}
 E_{\rm int}=\frac{\sigma}{2}\int {d^3x}~ a_k (\vec{x}) {\gamma_k(\vec{x})}. 
 \ee 
Formula (\ref{v-interaction}) obviously implies that two parallel infinitely long vortices repel each other, in contrast with magnetism when the parallel wires with   the currents in the same direction attract each other, though the formal expression for magnetic case is identically the same as  (\ref{v-interaction}).
The difference is of course due to the law of induction when the electric field is induced under a  small variation of the system.  
Our superfluid vortices are not subject to the law of induction, and therefore they repel.
  
 Once the vector potential $a_k (\vec{x})$ is known, the  corresponding physical velocity field $g_k(\vec{x})$ can be computed from 
 (\ref{a-potential})  as usual. Therefore, for any 
distribution of vortices the corresponding  classical  velocity field can be explicitly computed.  
However, our vortices are complicated  objects, and simplified  description of infinitely long straight vortices  is not quite appropriate description for questions we want to address. Nevertheless, the algebraic and topological structure of the fields (\ref{analogy}), (\ref{a-potential}) we introduced in this section will play an important role  in the  analysis which follows. 

 The correlation function $\la I^2\ra$   which enters the fundamental relation (\ref{n_s})    is proportional to  the   following correlation function  (up to some normalization factors)
  \be
  \label{W_new1}
 \la W_k, W_k\ra = \lim_{k\rightarrow 0}\int e^{i\vec{k}\cdot(\vec{x}-\vec{x}')}\frac{dS_k}{2\pi}   \frac{dS_k'}{2\pi}  \la  \gamma_k(\vec{x}),\gamma_k(\vec{x}')\ra ~~~~. 
  \ee
  The corresponding expectation value $\la W^2\ra$ does not depend on the position of the slice $dS_k$ along $k$ direction as a result of 
 exact property $\partial_k \gamma_k(\vec{x})=0$ as explained above. It implies that $\la I^2\ra$ also holds this feature. 
    
 \section{\label{susceptibility}Superfluidity and the winding number susceptibility $\la I^2\ra$.}
  The main goal of the present section is to derive and analyze the effective action which  accounts 
  for the complicated dynamics of the   vortices introduced  above.  Using the corresponding action we compute   the expectation value of the winding susceptibility $\la I^2\ra$, which is the main object of our studies. A hope is that these computations  will provide us with a hint on nature of   configurations which destroy the superfluidity.   Therefore, the  corresponding vortex configurations which are responsible for destroying the superfluidity can be thought as the configurations which control  the phase transition at $T_c$. 
  
   The technique which will be used in the present studies is well established one, and it  is commonly known to particles physics and condensed matter communities.  
   For convenience of the readers we review this technique in  Appendix \ref{deformed} which was previously developed for a   QCD-like model with  the main purpose to emphasize on some amazing similarities between the two systems. In particular, we show how the mass gap is generated in both systems as a result of mixture of the Goldstone mode with an  auxiliary topological field.
   
   We should warn a reader   from the  start  that we have made a large number of assumptions and approximations in course of our analysis in this section. In particular, it is quite obvious that the dynamics of the vortices, accounting for their numerous configurations with different densities and topologies is very important  for computations of $\la I^2\ra$. Such a computation is obviously a very ambitious goal which is well beyond the scope of the present work.
   We shall specifically formulate those assumptions along the course of our presentation. However, we believe that the basic consequences of this framework   are quite generic and robust  and not very sensitive to those approximations, at least for the computation of the $\la I^2\ra$ which is the main subject of the present work.

  The organization of this section is as follows. In    subsection \ref{auxiliary}   we     introduce  the auxiliary non-propagating topological field and derive the corresponding effective action, similar to  the Chern Simons term which is a conventional technique in studies of  the topologically ordered phases. 
  
  Using this topological action   we compute in subsection \ref{winding}  the winding susceptibility  $\la I^2\ra$ and analyze its properties. The main feature of this correlation function  is a generation  of the mass gap $\Delta_{\eta}(T)$ as shown in Section \ref{mass}.
  The obtained expression for the gap  is highly sensitive to superfluid density: it vanishes at the critical temperature $T=T_c$ when $n_S=0$.  This unique feature obviously implies that this gap  cannot be identified with the roton's excitations which are present in the system above and below the critical temperature. Therefore, we dubbed the corresponding excitations as vortons, which is the subject of Section  \ref{numerics}. 
  
  One should comment here  that the obtained  behaviour of $\la I^2\ra$  is very similar to known properties of $\la Q^2\ra$   
  which plays a crucial role in resolution of the so-called $U(1)_A$ problem   in QCD. In both cases the mass is generated as a result of  mixing of the Goldstone mode  with a non-propagating topological auxiliary field. In weakly coupled ``deformed QCD" all claims  can be made very precise as  reviewed    in  Appendix \ref{deformed} for convenience.

    \subsection{\label{auxiliary}Superfluidity and auxiliary gauge fields}

  The first step to proceed with this program  is to construct the current density $ j_k(\mathbf{x})$  for a generic configuration of the $n^{(i)}$ vortices (\ref{vorticity3}) directed along   $k-$th component   at point $\mathbf{x}^{(i)}_{\perp}$ as expressed below
  \be
  \label{current}
  j_k(\mathbf{x})=\sum_i 2\pi n^{(i)} \delta^2_k \left(\mathbf{x}_{\perp}-\mathbf{x}^{(i)}_{\perp}(x_k)\right),
  \ee
  where the position of the $i$-th vortex $\mathbf{x}^{(i)}_{\perp}(x_k)$ obviously depends on $x_k$ coordinate.  In this formula we obviously ignored the internal structure of the vortices by assuming that  the vortices  are infinitely thin. 
  Furthermore, we treat the vortices as heavy semiclassical objects, which fluctuate only slowly. In other words, we treat the vortices as the  static classical objects, which is obviously   not a good approximation to study the real time dynamics of the vortex network when the temperature approaches $T_c$ from below. Our justification for this ``static" approximation is based on observation that a similar assumption in a static-like   ``deformed QCD" (see  Appendix \ref{deformed}) and strongly coupled QCD produce   identically the same algebraic structure in computations of the   topological susceptibility  $\la Q^2\ra$. Our conjecture is that  a similar  feature of rigidity and stiffness   holds  in computations of the topological characteristic, the $\la I^2\ra$, in the  superfluid systems as well.

  The current  (\ref{current}) is analogous  in all respects to the topological density operator in ``deformed QCD" model  describing a  generic configuration of the point-like static monopoles  with arbitrary positions and colour orientations, see  Appendix \ref{deformed} for the details. Our goal here is to construct the partition function ${\cal{Z}}$ by integrating over all possible configurations (\ref{current}), similar to integrating over all possible monopole's configurations in ``deformed QCD" model, see Appendix \ref{deformed}.  
     
  Our next step is to  insert  the  delta function  into the path  integral with the field $b_i(\mathbf{x})$ acting as a Lagrange multiplier
\be
\label{delta}
&&\delta  \Big(   j_k(\mathbf{x})-   \frac{ \big[\epsilon_{ijk} \partial_{i}  g_j (\mathbf{x}) \big]}{2\pi} \Big) \nonumber\\
&&\sim\int {\cal{D}}[b_i]e^{ i\int d^3x ~b_k(\mathbf{x})\cdot \big(   j_k(\mathbf{x})-  \frac{1}{2\pi}    \big[\epsilon_{ijk} \partial_{i}  g_j (\mathbf{x}) \big]\big)}, 
      \ee
     where  $ j_k(\mathbf{x})$ in this formula is treated as the original expression (\ref{current}) for the density circulation    expression 
		including the fast velocity field, while $\epsilon_{ijk} \partial_{i}  g_j (\mathbf{x}) $ is treated as a slow-varying   external source describing the long  distance physics, similar to the treatment in Appendix \ref{deformed} of the auxiliary non-propagating topological fields in QCD.
		Furthermore, our formula  (\ref{delta}) corresponds to the normalization when  the current  $j_k(\mathbf{x})$ carryes integer fluxes of circulation.  
		
		Our task now is to integrate out the original fluctuating vortices with different shapes (treated here as the  fast degrees of freedom) and describe the large distance physics in terms of effective slow varying fields represented by 	the Lagrange multiplier 	$b_i(\mathbf{x})$ and 
	the auxiliary 	 $g_i (\mathbf{x})$ field. One should emphasize that  $g_i (\mathbf{x})$ field in formula (\ref{delta}) should not be confused 
	with local  velocity field $\vec{g}(\mathbf{x})=m\vec{v}$ in our previous discussions. The difference is that  $\big( b_k(\mathbf{x}), g_i (\mathbf{x})\big)$ as well as 
	$\gamma_k(\mathbf{x})=\left[\epsilon_{ijk} \partial_{i}  g_j (\mathbf{x}) \right]$ fields  entering  formula (\ref{delta}) will effectively describe the long range behaviour of the system after the 
	integrating  out the original fast fluctuating vortices with different shapes and positions, while the original $g_i (\mathbf{x})$ field describes a specific fast fluctuating local velocity distribution for a given vortex configuration. We keep the same notations because in what follows we will dealing exclusively with long-ranged effective $\big[ b_k(\mathbf{x}), g_i (\mathbf{x})\big]$  auxiliary fields.  
	
	To construct the partition function $\cal{Z}$  one should integrate out the original large  fluctuating vortices with different shapes, which is obviously  a very hard technical problem. This task  looks even harder because we should do these computations in the background of long range  $\big[ b_k(\mathbf{x}), g_i (\mathbf{x})\big]$ auxiliary fields. A similar,  but  technically   less complicated  problem (when one  should integrate over all positions and colour orientations of the monopoles) has been  explicitly carried out  in  the so-called ``deformed QCD" model, see Appendix \ref{deformed} for the details and comparison with  present analysis. One part of the effective description, the topological term $\beta H_{\rm top}\big[ b_k(\mathbf{x}), g_i (\mathbf{x})\big]$,   can be established  immediately from (\ref{delta}) as it represents the topological portion of the energy  and it is expressed exclusively in terms of auxiliary long ranged auxiliary  fields,  
	\be
	\label{Z1}
	&&{\cal{Z}} \sim \int {\cal{D}}[b] {\cal{D}}[g]{\cal{D}}[\alpha]e^{-\beta H[b,g,\alpha]- \beta H_{\rm top}[b, g]}\\
	&& \beta H_{\rm top}[b, g]=  \frac{i}{2\pi}  \int_{\mathbb{M}}  d^{3}x     b_k(\mathbf{x})  \big[\epsilon_{ijk} \partial_{i}  g_j (\mathbf{x}) \big]. \nonumber
			\ee
	The computations of $H[b,g,\alpha]$ is much more complicated task as we already mentioned. We should remark here that    $\alpha$ field  is included  into the effective formula for  $H[b,g,\alpha]$ because it describes   the long range light  Goldstone field  (\ref{G}) which must remain in the effective description 
 after integrating out all heavy degrees of freedom and complicated net of vortices. 
  	As explained above, in a class of simple models such kind of computations can be explicitly carried out and compared  with independent exact results as discussed in Appendix \ref{deformed}. Our present case obviously does not belong to this  class (of simple models) where exact computations are feasible. However we can establish the structure of  $H[b,g,\alpha]$ from  symmetry reasoning. 
	
	The argument goes as follows. The current (\ref{current}) which describes density of circulation as discussed after equations (\ref{vorticity2}), (\ref{vorticity3}) is exactly conserved for any closed vortices. This conservation is manifestation  of formal property of the system that the winding number remans the same for any slice along any direction. The corresponding current expressed in terms of the auxiliary long range field is also conserved, 
	$\partial_k[\epsilon_{ijk} \partial_{i}g_{j} (\mathbf{x})]=0$. Therefore, the Lagrange multiplier
$b_k(\mathbf{x})$ field is in fact a ``gauge" auxiliary field  such that the ``gauge  transformation"
\be
\label{gauge}
b_k(\mathbf{x})\rightarrow b_k(\mathbf{x})-\partial_k \Lambda(\mathbf{x})
\ee
leave the partition function  (\ref{delta})  due to large vortices	unaltered. This ``gauge invariance" unambiguously fixes a possible structure of the long distance portion of energy  $H[b,g,\alpha]$ entering (\ref{Z1}). Indeed, the auxiliary ``gauge" field may enter the Gross-Pitaevskii Lagrangian (\ref{GP}) only in combination 
\be
\label{GP3}
 {\nabla_i}\Phi (\mathbf{x}) \rightarrow \left[ {\nabla_i}-i{b_i}(\mathbf{x})\right]\Phi (\mathbf{x}).
\ee
	   After integrating out all the heavy degrees of freedom this ``gauge shift" implies that the relevant Hamiltonian describing the Goldstone field (\ref{G})
	   assumes the following  form
	   \be
	   \label{H_G}
	H_G =    \frac{n_S}{2m} \int_{\mathbb{M}} {d^3x}\big[{\partial_i}\alpha(\mathbf{x})-{b_i}(\mathbf{x})\big]^2 +{\rm interactions},~~~~~
	   \ee
	   where we neglected higher order corrections in fields and derivatives which follow from the expansion of the  Gross-Pitaevskii Lagrangian (\ref{GP}).    
	   
	   While the local interaction terms can be neglected for the qualitative analysis,  the long-ranged   term which is similar to the instantaneous coupling in electrodynamics which has the form (\ref{v-interaction}) must be included into the consideration. The simplest way to account for this  long-ranged interaction is to represent it in terms of the auxiliary vector potential 
	  $ a_k (\mathbf{x})$  which is related to the auxiliary $g_j (\mathbf{x})$ field precisely in the same way as original fast fluctuating fields   as given by eq.(\ref{a-potential}). Therefore, the interaction term (\ref{v-interaction}) can be written as follows
	  	    \be
 \label{H_int}
 H_{\rm int}=-\frac{\sigma}{2} \int_{\mathbb{M}} {d^3x}~ a_k(\mathbf{x}) \vec{\nabla}^2 a_k (\mathbf{x}). 
 \ee  
Similarly, the topological term (\ref{Z1}) can be also rewritten in terms of the ``gauge" field 	  $a_k(\mathbf{x}) $ as follows
	   \be
	   \label{S_top}
S_{\rm top}[b, a]=	   -\frac{i}{2\pi}  \int_{\mathbb{M}}  d^{3}x     b_k(\mathbf{x}) \vec{\nabla}^2 a_k (\mathbf{x}),  
\ee
where we always assume the ``Coulomb gauge" $\partial_k a_k(\mathbf{x}) =0$ for the auxiliary $a_k(\mathbf{x}) $ gauge field.

We are now in position to collect all relevant terms and represent the partition function to be used in following sections as follows
\be
	\label{Z-final}
	&&{\cal{Z}} \sim \int {\cal{D}}[b] {\cal{D}}[a]{\cal{D}}[\alpha]e^{-\beta \big(H_G[b,\alpha]+H_{\rm int}[a] + H_{\rm top}[b, a]\big)}\nonumber \\
	&& \beta H_{\rm top}[b, a]=   -\frac{i}{2\pi}  \int_{\mathbb{M}}  d^{3}x     b_k(\mathbf{x}) \vec{\nabla}^2 a_k (\mathbf{x}), 	\\
	&&	H_{\rm int}[a]=-\frac{\sigma}{2} \int_{\mathbb{M}} {d^3x}~ a_k(\mathbf{x}) \vec{\nabla}^2 a_k (\mathbf{x}), \nonumber \\
	&& H_G[b, \alpha]=    \frac{n_S}{2m} \int_{\mathbb{M}} {d^3x}\big[{\partial_i}\alpha(\mathbf{x})-{b_i}(\mathbf{x})\big]^2.\nonumber
		\ee
Few comments are in order. First of all, in formula (\ref{Z-final}) we neglected the  dynamics of the vortices,  including their mutual interactions, their numerous configurations with different densities, shapes and topologies, their    internal structure, etc. We also ignored all types of interactions (except for the crucial instantaneous inter-vortex interaction term $\sim \sigma$) as we keep only the dominant linear and quadratic terms. 

This procedure    allows us to  compute  the path integral exactly as the  relevant integrals have simple  Gaussian structure. Essentially, we treat the vortices as the static objects as explained after eq.(\ref{current}). 
The accounting  of the corresponding  dynamics   is obviously a very ambitious goal which is well beyond the scope of the present work. In principle, the corresponding computations can be carried out using the ``vortex-ring model"  \cite{Williams},  which basically employs the    renormalization group technique, or explicit numerical simulations \cite{Shenoy}. 
 However, we believe that all these complications may change  the numerical  parameters which enter the expression for  $\la I^2\ra$ as discussed in Section \ref{mass}. However, they cannot drastically change the basic algebraic structure and the basic features  
 of  the topologically rigid correlation function (\ref{I_final}) which represents  the main focus of the present analysis.

Our next comment is as follows. We   treated the Lagrange multiplier field $b_k(\mathbf{x})$ in eq.(\ref{Z-final}) as auxiliary, non-propagating  gauge potential. In other words, we assume that the gauge invariant combination $\big(\partial_ib_j(\mathbf{x})-\partial_jb_i(\mathbf{x})\big)$, which is allowed by the symmetry, nevertheless  cannot be generated as a result of any interactions.

Our next remark is
 as follows: the topological action with imaginary $i$ which enters (\ref{Z-final}) should not be considered as a  signal of  violation of unitarity.
 These fields do not have conventional kinetic terms, and do not propagate.  
 The fields $[b_k(\mathbf{x}), a_k(\mathbf{x})]$ should be treated as an auxiliary fields saturating the path integral, similar to  computation of a conventional integral using a steepest descent approximation when a saddle point lies in a complex plain outside the region of definition of the original variables. Analogous  effect also occurs in ``deformed QCD" model reviewed in Appendix \ref{deformed}, where similar  ``imaginary" action (\ref{eta-action})  reproduces the exact correlation functions  obtained by a different independent approach. 
 
 Our final comment is about the ``gauge invariance" expressed by eqs.(\ref{gauge}),(\ref{GP3}). Similar structure has been discussed previously in the literature \cite{book}, \cite{Gorsky}. The difference is that the gauge vector potential in  \cite{book},  \cite{Gorsky} is essentially an external background field  representing the twisted boundary conditions. In contrast, our gauge field $b_k(\mathbf{x})$ is introduced as a Lagrange multiplier, and it   is the  fluctuating (though non-propagating), field which describes the dynamics of internal fluctuating vortices with variety of shapes and positions. In our framework one and the same  ``gauge field" $b_k(\mathbf{x})$ enters  not only the Goldstone portion  of the Hamiltonian, $H_G[b, \alpha]$, but  it  also enters   the topological portion of the Hamiltonian  $H_{\rm top}[b, a]$ according to eq.(\ref{Z-final}).

  \subsection{\label{winding} Winding number susceptibility $\la I^2\ra$}
  The main goal of this subsection is to compute the winding number susceptibility $\la I^2\ra$ defined by eq. (\ref{I2}) using the partition function derived in previous subsection and given by (\ref{Z-final}).  We follow the textbook \cite{book} and define the ``gauge invariant" winding number susceptibility  
    \be
  \label{I4}
&&\la I_k,  I_k\ra =   \lim_{\mathbf{k}\rightarrow 0}\int_{\mathbb{M}}  {d^3x} \int_{\mathbb{M}}e^{i\mathbf{k}\cdot(\mathbf{x}-\mathbf{x}')}    {d^3x'}  \\ 
 &&\cdot\la  \big({\partial_k}\alpha(\mathbf{x})-b_k(\mathbf{x})\big), \big({\partial_k}\alpha(\mathbf{x}')-b_k(\mathbf{x}')\big)\ra .  \nonumber
  \ee
  which respects the symmetry (\ref{gauge}). As explained above this ``gauge invariance"  is a consequence of the ``current conservation"
  $\partial_k\gamma_k(\mathbf{x})=0$. 
  The expectation value   (\ref{I4}) should be computed using the  partition function defined by (\ref{Z-final}). The corresponding calculations can be carried out exactly because the Hamiltonian is quadratic in our approximate treatment.
  
 To proceed with computations, we first integrate out the auxiliary $b_k(\mathbf{x})$ field which enters (\ref{Z-final}) as a Lagrange multiplier.    It gives the constraint-like relation: 
  \be
  \label{b}
 \big[ b_k(\mathbf{x})- {\partial_k}\alpha(\mathbf{x})\big]=i\left(\frac{mT}{n_s}\right)\frac{\vec{\nabla}^2 a_k(\mathbf{x})}{2\pi}.
  \ee
  Now the susceptibility is expressed exclusively in terms of the topological $a_k(\mathbf{x})$ field: 
   \be
  \label{I5}
 &&\la I_k,  I_k\ra =  \lim_{\mathbf{k}\rightarrow 0}\int_{\mathbb{M}}  {d^3x} \int_{\mathbb{M}}   {d^3x'} e^{i\mathbf{k}\cdot(\mathbf{x}-\mathbf{x}')}  
{\cal{I}}({a_k})  \\  
 &&{\cal{I}}({a_k})=-  \left(\frac{mT}{n_s}\right)^2 
   \la  \frac{\vec{\nabla}^2 a_k(\mathbf{x})}{2\pi},\frac{\vec{\nabla}^2 a_k(\mathbf{x'})}{2\pi} \ra.  \nonumber\\
 \nonumber
    \ee
   Our next step is   to eliminate the  Lagrange multiplier  $b_k(\mathbf{x})$ in the Hamiltonian (\ref{Z-final}) using constraint  (\ref{b}). The corresponding simple algebraic manipulations lead to the following quadratic form for the Hamiltonian $H_{\rm tot} [a, \alpha]$ 
 \be
 \label{H}
&& \frac{H_{\rm tot} [a, \alpha]}{T} = \left(\frac{mT}{2n_s}\right) \cdot \int_{\mathbb{M}}  d^{3}x   \left(\frac{\vec{\nabla}^2 a_k(\mathbf{x})}{2\pi}\right)^2  \\
 &-& \int_{\mathbb{M}} {d^3x}\left[\frac{\sigma}{2T} a_k(\mathbf{x}) \vec{\nabla}^2 a_k (\mathbf{x})+\frac{i}{2\pi} \partial_k\alpha(\mathbf{x}) \vec{\nabla}^2 a_k (\mathbf{x})\right].\nonumber
 \ee 
  To complete the computations of the correlation function (\ref{I5}) one should diagonalize the quadratic form (\ref{H}) using conventional trick to shift  the variables from $ a_k(\mathbf{x})$ to  $a'_k(\mathbf{x})$:
  \be
  \label{shift}
   \left(\frac{\vec{\nabla}^2 a_k(\mathbf{x})}{2\pi}\right)= \left(\frac{\vec{\nabla}^2 a'_k(\mathbf{x})}{2\pi}\right)+i \left(\frac{n_s}{mT}\right)\partial_k\alpha(\mathbf{x}). 
  \ee
  This change of variables brings the Hamiltonian (\ref{H}) to the desired diagonal form
   \be
 \label{H1}
&& \frac{H_{\rm tot} [a, \alpha]}{T} = \left(\frac{mT}{2n_s}\right) \cdot \int_{\mathbb{M}}  d^{3}x   \left(\frac{\vec{\nabla}^2 a'_k(\mathbf{x})}{2\pi}\right)^2  \\
 &+& \int_{\mathbb{M}} {d^3x}\left[ \left(\frac{n_s}{2mT}\right) \big(\partial_k\alpha (\mathbf{x})\big)^2- \frac{\sigma}{2T} a'_k(\mathbf{x}) \vec{\nabla}^2 a'_k (\mathbf{x})\right].\nonumber
 \ee 
  
  Before we proceed with computations we want to make few technical comments. First, we neglected the singular term $\sim \partial_k\alpha (\mathbf{x})\frac{1}{\vec{\nabla}^2} \partial_k\alpha (\mathbf{x})$ which formally emerges  from vortex-vortex interaction $\sim \sigma$. Naively, it could be    interpreted as a generation of the mass for the Goldstone field $\sim \alpha^2$ if we formally cancel the singularity $\partial_k\partial_k
  \frac{1}{\vec{\nabla}^2}\sim 1$. Such a   manipulation with singular operator formally leading to $\alpha^2$ term    is   obviously incorrect. The singularity   emerges as a result of  our ignorance of the vortex stricture at small distances. Simple way to account for this structure is to replace $\frac{1}{\vec{\nabla}^2}\rightarrow \frac{1}{\vec{\nabla}^2-\mu^2}$ with some typical mass scale $\mu$. This cutoff procedure obviously leads to a desired result that the Goldstone field enters the Hamiltonian only through the derivative 
  $\sim \partial_k\alpha (\mathbf{x})$ without violating any fundamental theorems. In deriving (\ref{H1}) we also neglected the terms which can be integrated by parts. Finally, we   dropped  the terms which  vanish as a result of the ``Coulomb gauge" choice, $ \sim \partial_ka_k (\mathbf{x})=0$.
  
  Now we are in position to proceed with computations of the susceptibility $\la I^2\ra$ defined by (\ref{I4}), (\ref{I5}).
  We express the original fields  $ a_k(\mathbf{x})$   in terms of the shifted variables  (\ref{shift}) to arrive to
  \be
  \label{I6}
  &&\la I^2\ra =   \lim_{\mathbf{k}\rightarrow 0}\int_{\mathbb{M}}  {d^3x} \int_{\mathbb{M}}e^{i\mathbf{k}\cdot(\mathbf{x}-\mathbf{x}')}    {d^3x'}  
  \big[{\cal{I}({\alpha})}+{\cal{I}}({a'_k})\big]  \nonumber \\
&&{\cal{I}({\alpha})}=  \la   {\partial_k}\alpha(\mathbf{x}) ,  {\partial_k}\alpha(\mathbf{x}') \ra    \\
 &&{\cal{I}}({a_k})=-  \left(\frac{mT}{n_s}\right)^2 
   \la  \frac{\vec{\nabla}^2 a'_k(\mathbf{x})}{2\pi},\frac{\vec{\nabla}^2 a'_k(\mathbf{x'})}{2\pi} \ra ,   \nonumber
    \ee
    where the corresponding expectation values must be computed using   Hamiltonian (\ref{H1}). 
  The computations are straightforward because the 
  Hamiltonian (\ref{H1}) is quadratic. From now on we drop  the prime-sign in ${\cal{I}}({a_k})$ to simplify notations.
  
  It is very instructive to compute ${\cal{I}({\alpha})}$ separately by first  ignoring  the   term ${\cal{I}}({a'_k}) $ related to auxiliary gauge fields.
  The result is 
  \be
  \label{I7}
  {\cal{I}({\alpha})}=  \frac{mT}{n_s}\cdot \left(\frac{\delta(x-x')}{L_2L_3}+\frac{\delta(y-y')}{L_1L_3}+\frac{\delta(z-z')}{L_2L_1}\right),~~~~
  \ee
  where the first term is computed assuming the path $\Gamma\in \mathbb{M}$ goes along $\hat{x}$ direction, while the second and third terms 
  are generated due to the paths along $\hat{y}$ and $\hat{z}$ directions correspondingly as explained after eq. (\ref{I2}). 
  Indeed, the correlation function ${\cal{I}({\alpha})}$ obviously vanishes at  $ \mathbf{x}\neq \mathbf{x}'$  as $\vec{\nabla}^2\frac{1}{|\mathbf{x}-\mathbf{x}'|}=0$ at 
 $ \mathbf{x}\neq \mathbf{x}'$ in infinite volume. To get a better idea about the structure  of the singularity for finite manifold $\mathbb{M}$ when a singular limit is approached along a specific  $\hat{x}$ direction it is convenient to approximate the relevant portion of the Hamiltonian (\ref{H1}) as 
  $\int dx\left(\frac{n_sL_2L_3}{2mT}\right) \big(\partial_x\alpha ({x})\big)^2$. This Hamiltonian  precisely generates the first singular term in eq. (\ref{I7}). Two other terms can be obtained in a similar manner.

  The result   (\ref{I7})  is  well anticipated  contact term. The corresponding contribution to $\la I^2\ra_{\alpha}$ after integration over volume $ \int {d^3x} {d^3x'} $ is
   \be
   \label{I8}
   \la I^2\ra_{\alpha}=  \int   {d^3x} \int  {d^3x'}  \cdot  {\cal{I}({\alpha})}= \left(\frac{3mT}{n_s}\right)V.
   \ee
  Few comments are in order. First of all, if we substitute (\ref{I8}) to our original formula  (\ref{n_s}) for $n_s$ we get identity $n_s=n_s$,
  which is   expected result as we neglected many important elements, including ${\cal{I}}({a_k})$ term, and all the interaction terms. We also neglected the internal vortex structure, interaction of the Goldstone field with the vortices, etc, which in principle can be recovered from the original Gross-Pitaevskii Lagrangian (\ref{GP}). Nevertheless, this ``trivial" result shows that our formal manipulations with the path integral in sections \ref{auxiliary}, \ref{winding}, including operations with  auxiliary topological fields $(b_k(\mathbf{x}), a_k(\mathbf{x}))$, are consistent with all general principles. Furthermore, these  formal  manipulations lead to the correct normalization (\ref{I8}) which we consider as a highly nontrivial consistency check of our path integral approach when the main simplifications (vortices are infinitely thin and static) do not lead to any fundamental contradictions. We expect that while the  $\la I^2\ra_{\alpha}$ term reproduces the basic normalization, 
  the accounting for auxiliary topological fields  (effectively describing the dynamics of vortices) and  interactions between them and the Goldstone field should, in principle, generate a nontrivial equation relating $n_s$ and input parameters of the theory.  Furthermore, it is quite obvious that 
  the dynamics of the vortices, accounting for their numerous configurations with different densities (affected by the temperature), will also drastically influence the numerical analysis. 
  
 Such a computation  is obviously a very ambitious goal which is well beyond the scope of the present work. In fact, we do not even attempt to go along this direction as we keep only the quadratic terms in the Hamiltonian, ignoring all the interaction terms, except for the instantaneous long range vortex-vortex interaction proportional to $\sigma$. 
 
 Our goal of the present work is much more modest: we want to demonstrate the emergence of the two basic features 
 in the system. First, in spite  of the simplifications we have made,  the correlation function $\la I^2\ra$ exhibits   the   algebraic structure which is identically the same as   $\la Q^2\ra$ in QCD.
 It shows that some ``topologically rigid" correlation functions are not very sensitive to those simplifications.
 Secondly, our formal procedure leads to a generation of the mass scale, resulting from the auxiliary topological fields. The generation of this scale can be seen already at the lowest quadratic level as we argue below in section \ref{mass}. 
 
  We expect that accounting for the dynamics and the interacting terms using some simplified models such as ``vortex ring model" from refs. \cite{Williams, Shenoy} may renormalize  some numerical parameters for the correlation function $\la I^2\ra$. 
 However, we do not expect  that the accounting for the  interaction terms may drastically  change the basic algebraic structure (\ref{I_final}) to be discussed below.   The main argument behind  this assert  is that the  topological susceptibility  $\la Q^2\ra$ in QCD shows a  similar  ``topological rigid" feature  when real strongly coupled QCD is replaced by the weakly coupled  ``deformed QCD" model as discussed in Appendix \ref{deformed}.  
   
   \subsection{\label{mass}  Generation of the mass scale}
  With these comments in mind  from previous subsection we now  proceed   with computations of ${\cal{I}}({a_k}) $ contribution defined by (\ref{I6}). The relevant for these computations part of the Hamiltonian (\ref{H1}) 
   includes four- derivative  term which requires some extra care. We present  $a'_k(\mathbf{x})$-dependent term in the Hamiltonian as follows
   \be
   \label{4-derivatives}
     \frac{H_{\rm tot} [a]}{T} &=& \left(\frac{mT}{2n_s}\right)  \int   d^{3}x    \frac{a'_k(\mathbf{x})}{2\pi}\big[\vec{\nabla}^2\vec{\nabla}^2-\Delta_{\eta}^2\vec{\nabla}^2 \big]\frac{a'_k(\mathbf{x})}{2\pi},\nonumber  \\
  \Delta_{\eta}^2&\equiv& (2\pi)^2\left(\frac{\sigma n_s}{m T^2}\right)=\eta\left(\frac{2\pi n_s}{m T}\right)^2,
   \ee
  where in the last line we expressed the mass parameter $\Delta_{\eta}$ in terms of the    dimensionless parameter $\eta\sim1 $ replacing the dimensional  parameter $\sigma=\eta (n_s/m)$ effectively describing the strength of the vortex-vortex interactions,  as discussed after eq. (\ref{v-interaction}). One should comment here that a similar structure with  four derivatives  also occurs  in ``deformed QCD" model, see (\ref{S_QCD}).
  Therefore, we have some experience  how to proceed with computations.    
  
  To evaluate  the correlation function  ${\cal{I}}({a_k}) $ defined  by eq.(\ref{I6}) we have to find the inverse of the 4-derivative  operator entering (\ref{4-derivatives}), i.e.
  $\big[\vec{\nabla}^2\vec{\nabla}^2-\Delta_{\eta}^2\vec{\nabla}^2 \big]^{-1}$.
  To proceed with this task,  we use a standard  trick to represent the 4-th order operator $\left[  \vec{\nabla}^2 \vec{\nabla}^2- \Delta_{\eta}^2  \vec{\nabla}^2 \right]$  
    as a combination of  two terms with the opposite signs.   To be more specific, we write 
\be
\label{inverse1}
\frac{1}{\left[  \vec{\nabla}^2 \vec{\nabla}^2- \Delta_{\eta}^2 \vec{\nabla}^2 \right]}= \frac{1}{\Delta_{\eta}^2}\left(\frac{1}{  \vec{\nabla}^2 - \Delta_{\eta}^2 } -  \frac{1}{  \vec{\nabla}^2 } \right) , 
\ee
such that the Green's function for $a_k'(\mathbf{x})$ field which enters the expression for the winding susceptibility (\ref{I6}) can be represented as a combination of two Green's functions: conventional  massive field   and the ``ghost-like" field with ``wrong" sign. 
  Naively, the presence of 4-th order operator in eq. (\ref{4-derivatives})  is a signal that the ``ghost-like" instability develops   with violation of the unitarity and causality    occurring in the system. 
  However, this naive suspicious is obviously incorrect as the Hamiltonian (\ref{4-derivatives}) was derived from the original perfectly defined and unitary QFT. Technically, the absence of any fundamental deficiencies in our system is  explained by the fact that   $a'_k(\mathbf{x})$ field is not a dynamical field with canonical kinetic term.
  Instead, this field is an auxiliary topological field which does  not propagate as a conventional dynamical degree of freedom.   
  
  A very similar 
  structure (\ref{inverse})  is known to occur in strongly coupled  QCD. The corresponding topological field  in QCD is known as the Veneziano ghost. Precisely this structure is the key element in resolution of the celebrated  $U(1)$ problem when the $U(1)$ singlet Goldstone field  receives its mass  as a result of mixing of would be Goldstone field with the topological Veneziano ghost. We elaborate on this very close  analogy in Appendix \ref{deformed} and also in next section  \ref{interpretation} with specific comments on similarities and differences between the the winding number susceptibility $\la I^2\ra$ in superfluid system  and topological number susceptibility $\la Q^2\ra$ in confined QCD. In fact our notation for $\Delta_{\eta}$ in eq. (\ref{4-derivatives}) is inspired by this amazing analogy.

 We now proceed with explicit computations.   The   relevant correlation function   which enters the expression for the winding  susceptibility (\ref{I6}) can be explicitly computed using expression (\ref{inverse1}) for inverse operator  as follows 
 \be\label{I_a}
  &&{\cal{I}}({a_k}) =-  \left(\frac{mT}{2\pi n_s}\right)^2 \frac{\int{\cal{D}}[ a']e^{- \frac{H [a]}{T}}\vec{\nabla}^2 a_k' (\mathbf{x}), \vec{\nabla}^2 a_k' (\mathbf{x'})}{\int{\cal{D}}[ a']e^{- \frac{H [a]}{T}}} \nonumber \\
&&=-3\left(\frac{mT}{n_s}\right) \int \frac{\dd^3p  }{\left(2\pi\right)^3} e^{i\mathbf{p}\cdot(\mathbf{x}-\mathbf{x}')} \frac{p^4}{\Delta_{\eta}^2}\left[ - \frac{1}{p^2+\Delta_{\eta}^2} + \frac{1}{p^2} \right] 
\nonumber \\
   &&=-3\left(\frac{mT}{n_s}\right)\int \frac{\dd^3p}{\left(2\pi\right)^3}  e^{i\mathbf{p}\cdot(\mathbf{x}-\mathbf{x}')} \left[ \frac{p^2}{p^2+\Delta_{\eta}^2}   \right] \nonumber\\
   &&= -3\left(\frac{mT}{n_s}\right) \left[ \delta(\mathbf{x}-\mathbf{x}')-\Delta_{\eta}^2\frac{e^{-\Delta_{\eta}|\mathbf{x}-\mathbf{x}'|}}{4\pi |\mathbf{x}-\mathbf{x}'|} \right],  
  \ee
  where coefficient $3$ in front is due to three different components $a_k' (\mathbf{x})$  equally contributing to ${\cal{I}}({a_k}) $. 
The corresponding contribution to $\la I^2\ra_{a}$ after integration over volume $ \int {d^3x} {d^3x'} $ is
   \be
   \label{I_final}
   &&\la I^2\ra_{a}=  \int   {d^3x} \int  {d^3x'}  \cdot  {\cal{I}}({a_k})= \\
   &&-3\left(\frac{mT}{n_s}\right) \int   {d^3x} \int  {d^3x'}   \left[ \delta(\mathbf{x}-\mathbf{x}')-\Delta_{\eta}^2\frac{e^{-\Delta_{\eta}|\mathbf{x}-\mathbf{x}'|}}{4\pi |\mathbf{x}-\mathbf{x}'|} \right] . \nonumber
   \ee
   We are now  in position to make few important comments on the obtained results. First of all, formula (\ref{I_a}) has the structure which is identical to (\ref{QCD_final}) derived for the ``deformed QCD" model, where independent computations support this formal manipulations with topological auxiliary fields. Therefore, we are quite confident in the formal  technique employed  in this section.     
   
  Now, the contact term proportional to the delta function in eq. (\ref{I_final}) exactly cancels the contact term (\ref{I8}) generated by  the Goldstone field such that the formula for $\la I^2\ra$ assumes the form
    \be
   \label{I_eta}
  && \la I^2\ra = \la I^2\ra_{\alpha}+ \la I^2\ra_{a} +{\rm (other~ terms)}\\
   &&=3\left(\frac{mT V }{n_s}\right)  \int_{R_0}    {d^3x}\left(\Delta_{\eta}^2\frac{e^{-\Delta_{\eta}r }}{4\pi r}\right)+{\rm (others)},\nonumber
      \ee
   where we introduced the cutoff parameter $R_0$ to parametrize the deviation for the  vortex-vortex interaction from the ideal case of infinitely thin vortices. The terms indicated  as  ``others" in eq. (\ref{I_eta})  account for all types of interactions which were consistently ignored in our simplified treatment  as we are  dealing with the quadratic terms in the Hamiltonian only.  We want to rewrite our expression (\ref{I_eta})     
   in the following  form which is quite convenient for our numerical estimates in next section \ref{numerics}, 
    \be
   \label{I_cutoff}
   &&\la I^2\ra =3\left(\frac{mT V }{n_s}\right)f(z) +{\rm (others)},   \\
    &&f(z)\equiv ~~ \int_{R_0} {d^3x}\left(\Delta_{\eta}^2\frac{e^{-\Delta_{\eta}r }}{4\pi r}\right), ~~ z\equiv \Delta_{\eta}R_0,  \nonumber
      \ee
  where dimensionless function $f(z=0)=1$ is normalized to one at $z=0$,   and becomes exponentially small   at large $z$.  One should emphasize that both dimensional parameters, $\Delta_{\eta}$ and cutoff scale $R_0$ (which represents an effective size of a vortex core) are obviously  a  nontrivial functions
  of density $n_s$ and temperature $T$, as one can see from definition (\ref{4-derivatives}). Therefore, $z$ is also a nontrivial parameter of the density $n_s$ and temperature. One can argue that $z$ in vicinity of the phase transition scales as $z\sim \tau^{-\beta/2}$, see eq. (\ref{gap_numerics}) for  the definition of the reduced temperature $\tau$. 
  
  The most important result of this subsection is an explicit  demonstration that the system generates the mass scale   as a result of vortex-vortex interaction. Such an interpretation of the obtained  gap (resulting from vortex-vortex interaction) is motivated   by the expression  (\ref{4-derivatives}) for  the mass $\Delta_{\eta}$ which  is explicitly  proportional to the strength of the vortex-vortex  interaction. It is naturally to assume that the corresponding scale $\Delta_{\eta}$ is related to some elementary excitations, see few comments on this assumption in next section. However, the corresponding excitation gap $\Delta_{\eta}$  cannot be identified (due to a number of reasons) with the well known  roton's  gap $\epsilon_0$ defined  by  eq.(\ref{epsilon}).   A simplest argument supporting this claim is that our gap $\Delta_{\eta}\sim n_s(T)$   vanishes at $T\rightarrow  T_c$ in contrast with roton's gap which remains finite in vicinity of the phase transition at $T\simeq T_c$ as reviewed in section \ref{eff_lagrangian}.

   \section{\label{numerics} The Vortons }
   In this section we want to elaborate on the nature of the gap $ \Delta_{\eta} $ and make few simple numerical estimates  for these new topological objects. They will be dubbed  the {\it vortons} in what follows. The corresponding gap in vicinity of the phase transition
      can be approximated  as follows
   \be
   \label{gap}
   \Delta_{\eta} (T\rightarrow T_c)\simeq \sqrt{\eta}\left(\frac{2\pi   n}{m T}\right)\cdot \left(\frac{T_c-T}{T_c}\right)^{2\beta}, ~ \beta=\frac{1}{2}, ~~~
   \ee
   where we assume that the critical exponent $\beta=1/2$ is as in conventional Landau-Ginsburg approximation\footnote{Experimentally $\beta\simeq 1/3$ rather than 1/2. The difference is normally explained using renormalization group procedure.   We consistently ignore all the interacting terms in our analysis. Therefore, we use ``bare value"  $\beta=1/2$ in our estimates.}. We will  provide some justification for the term  ``vorton"  later in the text. But first,  we want to explain some key properties of these unusual objects. 
 
   Most important feature of the vorton is its $T$- dependent gap (\ref{gap}).      Furthermore, this object completely disappears from the system 
   in normal phase at $T>T_c$ because it  exists in the system only when superfluid vortices exist. The vortons always accompany the vortices and disappear from the system together with the superfluid vortices at $T>T_c$.
  
  One can numerically estimate a typical gap for vortons in vicinity of the critical temperature when $T$ is close to $T_c$ as follows,  
   \be
   \label{gap_numerics}
   \Delta_{\eta} (T\rightarrow T_c) \sim 10 K\sqrt{\eta}\cdot \tau^{2\beta}, ~~ \tau\equiv \left(\frac{T_c-T}{T_c}\right). 
   \ee
  Numerically\footnote{We expect that these numerical estimates   change as a result of many effects mentioned at the end of subsection \ref{winding}, and which were completely ignored in present analysis. However, we do not expect that the fundamental properties of the system   may drastically change as a result of these simpifications.}, the gap $\Delta_{\eta}$   is very similar   to the roton's gap $\epsilon_0\simeq 8 K$ away from $T_c$. However,    the vorton's gap $\Delta_{\eta}$  becomes much smaller than the roton's gap at $(T_c- T)\ll T_c$ when the vortons  become light\footnote{It is interesting to note that a quasiparticle with such unusual dispersion relation (\ref{gap_numerics})  when the gap vanishes at $T=T_c$ has been postulated in \cite{Bobrov} and dubbed as ``helon". We cannot comment  on relation (if any)  between the two objects as the framework  and technique developed in sections \ref{winding}, \ref{mass} is drastically different from ref. \cite{Bobrov}.}.  
  
   If one assumes that dimensionless parameter $\eta$,  effectively describing the interaction between vortices (which themselves become ``fat"
   when the core of a vortex and its length start to  scale in a similar way  in vicinity of the phase transition \cite{Shenoy})    is proportional to the overlapping volume between two vortices than $\eta$ in vicinity of the phase transition scales as   $\eta\sim \tau^{-3\beta}$ while  the gap $\Delta_{\eta}\sim \tau^{ \beta/2}$. If one also assumes that the ``other" terms in (\ref{I_eta}) do not qualitatively change the physics, the winding number correlation function $\la I^2\ra$ vanishes at the critical temperature as $\la I^2\ra \sim \exp (-   \tau^{-\beta/2})$. These estimates should be taken with great precaution and  grain of salt  because  it is very hard to  justify  any of  these  assumptions.
     
  It is quite obvious that the vorton's excitations can be completely ignored far away from the phase transition. First, they become very heavy as one can see from $T$- dependent formula for the gap (\ref{gap}), and therefore,  they can be excited only in  a close  vicinity of $T\simeq T_c$.  In fact, this is  the main  reason for our approximate expression  (\ref{gap_numerics}). The second reason for an  additional suppression  of vortons  at low $T\ll T_c$ is the observation that the vortons are intimately related to vortices as they obviously contribute to the winding number correlation function $\la I^2\ra$ according to (\ref{I_eta}). Their contribution to other correlation function which does not include the vorticity operator is suppressed, see few comments below.  Therefore, when the network of the vortices is not yet formed, or not well- developed  at sufficiently low $T\ll T_c$ the vortons   are also absent in the system. 
  
  Another key characteristic of the vorton is its spin. Vorton is described in terms of the (axial) vector topological field $a_k(\mathbf{x}) $ which corresponds to  $S=1$.
  The coefficient $3$ in front of the correlation function (\ref{I_a}) can be interpreted as the degeneracy   factor $(2S+1)$. This coefficient $3$  is precisely what is required  for the  cancellation of the contact terms as expressed by eq.(\ref{I_eta}). 
   
 One may wonder: why and how the  topological field $a_k(\mathbf{x}) $ which was originally introduced into the partition function (\ref{Z1}), (\ref{Z-final})  as an auxiliary topological  vector   field  becomes a dynamical degree of freedom? The answer is that this auxiliary field  $a_k(\mathbf{x}) $ is mixing with the physical propagating Goldstone field $\alpha(\mathbf{x})$ in a course of diagonalization of the Hamiltonian (\ref{H1}) according to (\ref{shift}). Precisely this mixing   eventually makes  the auxiliary field  (some part of it) $a_k(\mathbf{x}) $ to become a  physical quasiparticle-like degree of freedom. At the same time, another portion   of the $a_k(\mathbf{x}) $ field remains a non-propagating degree of freedom and manifests itself as a contact term $\sim \delta(\mathbf{x}-\mathbf{x}')$ contributing to  the correlation function (\ref{I_final}) with the ``wrong sign" which is opposite to conventional contribution of the physical Goldstone field (\ref{I7}), (\ref{I8}). This ``wrong sign" provides a specific mechanism of removal  of the winding number   from the system which happens at the phase transition point $T_c$ when this removal is completed. 
 
 One can view formula (\ref{I_final})
 as an explicit manifestation of the ``dual" nature of the topological field $a_k(\mathbf{x}) $: the first   contact term  $\sim \delta(\mathbf{x}-\mathbf{x}')$ is obviously not related to any propagating degrees of freedom, while the second term is obviously related to a physical propagating degree of freedom, the vorton,  with a gap $\Delta_{\eta}$. 
 
 Such a ``dual" behaviour of the auxiliary topological field once  again amazingly resembles  the behaviour of the QCD Veneziano ghost  which contributes with a ``wrong" sign to the QCD topological susceptibility $\la Q^2 \ra$, mixes with the Goldstone field and becomes the  massive and propagating  $\eta'$ field. This resolution of the renowned  $U(1)$ problem as formulated by Witten and Veneziano in 1980 is well supported by a numerous  lattice simulations and  commonly  accepted by the QCD community,   see Appendix \ref{deformed} for details and references.  
 
 It is instructive to compute the correlation function $\la   {a'_i(\mathbf{x})} , {a'_j(\mathbf{x'})}  \ra$ itself for a better understanding of the nature of the topological auxiliary field. This computation  will also  provide a much closer analogy    with Veneziano ghost mentioned above. To accomplish this task we define the correlation function ${\cal{V}}_{ij}({a_k})$ which is directly related to the previously computed correlation function $ {\cal{I}}({a_k})$ as follows,  
  \be
 \label{V}
{\cal{V}}_{ij}({a_k})&=&-  \left(\frac{mT}{n_s}\right)^2 
   \la  \frac{  a'_i(\mathbf{x})}{2\pi},\frac{  a'_j(\mathbf{x'})}{2\pi} \ra ,\\
   {\cal{I}}({a_k})&=& {\vec{\nabla}_{\mathbf{x}}}^2 {\vec{\nabla}_{\mathbf{x}'}}^2 {\cal{V}}_{ii}({a_k}) \nonumber
 \ee
 where  ${\cal{I}}({a_k})$   is given by (\ref{I_a}) and (\ref{I_final}).  The computation of the correlation function ${\cal{V}}_{ij}({a_k})$ can be carried out in a similar manner with result
 \be
 \label{V_a}
  &&{\cal{V}}_{ij}({a_k}) =-  \left(\frac{mT}{2\pi n_s}\right)^2 \frac{\int{\cal{D}}[ a']e^{- \frac{H [a]}{T}}  a_i' (\mathbf{x}),  a_j' (\mathbf{x'})}{\int{\cal{D}}[ a']e^{- \frac{H [a]}{T}}}\\
&&= \delta_{ij}\left(\frac{mT}{\Delta_{\eta}^2n_s}\right) \int \frac{\dd^3p  }{\left(2\pi\right)^3} e^{i\mathbf{p}\cdot(\mathbf{x}-\mathbf{x}')}\left[  \frac{1}{p^2+\Delta_{\eta}^2} - \frac{1}{p^2} \right].  \nonumber 
  \ee 
The structure of this expression is very  suggestive. First of all,  the physical contribution with the gap $\sim \Delta_{\eta}$ is precisely the contribution of vortons in our previous computations of the winding number correlation function (\ref{I_a}) and (\ref{I_final}). 
The pole at $p^2=0$ with a ``wrong sign" which is present in (\ref{V_a}) might look very suspicious and dangerous. However, this pole is present in gauge-dependent correlation function  ${\cal{V}}_{ij}({a_k})$.    This pole generates the contact term $\delta(\mathbf{x}-\mathbf{x}')$ in gauge invariant correlation function   (\ref{I_a}) and (\ref{I_final}) which is obviously consistent with all fundamental principles of the theory. 

This pole does not contribute to any other correlation functions, with exception of  the winding number susceptibility $\la I^2 \ra$    where it manifests itself as the contact term. In other words,  the strange pole  does not correspond to any physical massless degrees of freedom as its only role is the  generation of    the contact term in the winding number correlation function  $\la I^2 \ra$, similar to QCD case when the Veneziano ghost contributes exclusively to $\la Q^2 \ra$ and nowhere else.  
Nevertheless, this pole at $p^2=0$ explicitly shows that the physics behind the contact term in gauge invariant correlation function   (\ref{I_a}) is due to the infrared (IR) rather than ultraviolet (UV) physics, which is very similar to QCD where one can show that the contact term in $\la Q^2 \ra$ is saturated by IR rather than by UV physics.

As we review  in Appendix \ref{deformed} the direct analog of eq.(\ref{V_a}) is eq. (\ref{A}) in ``deformed QCD" model. Furthermore, one can explicitly construct the physical Hilbert space in a such a way that the unphysical ghost-like particles are removed from the physical amplitudes  by imposing  the Gupta-Bleuler- like condition on the physical Hilbert space, see (\ref{gb}). We shall not elaborate on this topic in the present work, by mentioning that the pole in (\ref{V_a})  is really harmless with the ``only" trace that it contributes to the contact term with the ``wrong" sign and has the tendency to cancel the winding number which is present in the system.
This  tendency becomes a much more   profound phenomenon in vicinity of the phase transition when the original winding number vanishes at $T=T_c$ precisely as a result of this cancellation, and the system becomes a convention non-superfluid liquid at $T\geq T_c$.

 From these discussions it is quite obvious that the   manifestations of the vortons are  drastically different from conventional  observables resulted  from   fluctuations of  ``normal" degrees of freedom  such  the rotons and phonons. In particular, the vortons  cannot be observed as the pole in the well studied ``density-density" response function defined by (\ref{S}), (\ref{epsilon})  simply because the topological field $a_k(\mathbf{x}) $  does not couple directly to the density fluctuations. Instead, it couples to the circulations $\gamma_k(\mathbf{x}) $ as eq.(\ref{a-interaction}) states. It also explains why the vorton excitations do not manifest themselves in conventional inelastic X-ray and neutron scattering experiments. Rather, the vortons contribute to the winding number susceptibility ${\cal{I}}({a_k}) $ according to (\ref{I_a}). The interaction of the vortons  with other quantum fluctuations (which can be in principle computed using renormalization group procedure) should generate some contribution to the ``density-density" response function as well. However, we expect this contribution to be numerically strongly suppressed due to a small mixing of the topological field $a_k(\mathbf{x}) $ with the Goldstone field (\ref{shift}) at $T\simeq T_c$  when the vortons are sufficiently light (and therefore,   can be easily excited),  however  the superfluid density is already quite small $n_s\sim (T_c- T)^{2\beta}$ at $T\simeq T_c$. As we   mentioned previously, any effects of vortons at very low $T\ll T_c$ are exponentially small and can be safely ignored. 
 
 Our next  comment in this subsection is   devoted to the term ``vorton". We want to justify   our terminology by dubbing  the  topological excitations $a_k(\mathbf{x}) $ emerging in our framework as ``vortons". Originally, the vortons were introduced in particle physics (relativistic systems)   when a superconducting string may form a macroscopically large closed stable vortex loop.  The stabilization occurs due to condensation of another  field in the vortex core and non-dissipating current moving along the core of the vortex. Similar classical objects were also constructed in non-relativistic systems with Bose-Einstein Condensates, see \cite{Metlitski:2003gj} for references and details.  In particular, it has been shown in \cite{Metlitski:2003gj} that  the vorton carries the angular momentum, and the stability of the vortons can be understood in terms of conservation of angular momentum.    
 
 Our elementary objects with typical microscopical scales (\ref{gap_numerics}) are obviously quantum, not classical,    objects. Nevertheless, they have a number of common features which motivated   our terminology in the present work. 
 Indeed, the circulation field plays a key role in classical as well in quantum description of the vortons represented  by $a_k(\mathbf{x}) $ in our construction. Furthermore, the role of non-dissipating conserved current in construction of classical object  plays the circulation source $\gamma_k(\mathbf{x}) $ in present work. In addition, both objects, classical and quantum,  are characterized by angular momentum which is the spin $S=1$ in our construction of quantum vortons. Finally, the vortons in our framework explicitly contribute to the correlation function (\ref{I_a}), which is analogous to the computation of the integer winding number in construction of the classical configuration \cite{Metlitski:2003gj}. Based on these similarities one can view the quantum elementary vorton described   by $a_k(\mathbf{x}) $ in our construction 
 as a microscopical elementary ring of a circulating conserved current $\gamma_k(\mathbf{x}) $. These objects represent the   counterparts of the macroscopically large classical vortons. Such a view is, in fact, very close to Feynman's interpretation of relevant quasi-particles responsible for the breakdown of the superfluidity \cite{Feynman}.

 $\bullet$   We conclude this section with the following generic remark. 
    The main claim of this section is that the long range physics due to the topological configurations (in form of a complicated network of knotted, folded   and wrinkled vortices)  can be formulated in terms of an auxiliary topological fields 
    $[a_k(\mathbf{x}) , b_k(\mathbf{x})] $. These fields do not propagate. Rather, they effectively describe the long-range dynamics of a  complicated network of vortices. As such, these fields generate a macroscopically large contribution to the free energy expressed in terms  of $\la I^2\ra_a$. In a sense these effective fields are similar to instantaneous static Coulomb field which, of course, does not describe any  propagating physical photons, but nevertheless generates a macroscopically large static background.   In our framework the auxiliary  fields are approximately static, which is a direct    consequence of our assumption  formulated after eq. (\ref{current}). This assumption is somewhat supported {\it a posteriori} as these auxiliary topological fields indeed generate the dominant contributions     to $\la I^2\ra_a$ according to our computations in section \ref{susceptibility}.

    The topological fields $[a_k(\mathbf{x}) , b_k(\mathbf{x})] $, being the emergent gauge fields, do not completely decouple from the gauge invariant observables. In particular, they do    contribute to 
    the winding number correlation function $\la I^2\ra$ due to its  unique structure. Furthermore, $a_k(\mathbf{x})$    mixes with the Goldstone field which  becomes a massive propagating degree of freedom,   the vorton.
    The emergent mass scale  (\ref{gap}) manifests itself precisely by studying  this specific correlation function. 
    
    It is a highly nontrivial phenomenon, and it has its counterpart in relativistic particles physics, where a similar new scale is generated in a very much the same  way. This novel scale is parametrically suppressed $\sim 1/N$  in large $N$ QCD,  and it can be explicitly tested by studying  the   topological susceptibility correlation function,  $\la Q^2\ra$. Furthermore,  the Veneziano ghost which plays the same role as  auxiliary topological fields 
    $[a_k(\mathbf{x}) , b_k(\mathbf{x})] $ in our present construction, generates the contact term in  $\la Q^2\ra$.
    Precisely this contact term is the crucial element in the resolution of the celebrated  $U(1)$ problem.   The Veneziano ghost  does not contribute to any other correlation functions.   We review this well established mechanism   in a simplified ``deformed QCD" model in Appendix \ref{deformed}  with emphasize on analogy and similarities with topological features of the superfluid systems studied above.

  \section{\label{interpretation} QCD vs. Superfluidity:  $\la Q^2\ra$ vs. $\la I^2\ra$ } 
  In this section we want to discuss a number of similarities and differences between our computations  $\la I^2\ra$ for superfluid liquid and 
  the computations of $\la Q^2\ra$  in QCD. First, we list a number of formal similarities in subsection \ref{formal}. In subsection \ref{difference} we formulate the fundamental difference between the two systems. Finally, in subsection \ref{skeleton} we compare the long range structure observed in lattice simulations  in strongly coupled QCD  with vortex network studied   in Helium II experiments. 
  \subsection{Formal similarities} \label{formal}
  We already mentioned a number of formal similarities in correlation functions $\la I^2\ra$ for  superfluid system and $\la Q^2\ra$ for
  ``deformed QCD" model. We want to list them for convenience once again. The role  of eq.(\ref{Z-final}) plays (\ref{eta-action}) in ``deformed QCD" model. The equations  (\ref{I5}), (\ref{H}) in  superfluid system (which includes mixing with the Goldstone field)  are formally similar to (\ref{top_QCD}). Four derivative operator (\ref{4-derivatives})  in  superfluid system is identically the same as 4-derivative operator (\ref{S_QCD}) which occurs in ``deformed QCD" model. 
  Finally, $\la I^2\ra_a$  has the structure (\ref{I_a}) which is very much the same as $\la Q^2\ra$ given by eq.(\ref{QCD_final}) in the QCD like model.
  The  topological  auxiliary fields $[a_k(\mathbf{x}), b_k(\mathbf{x})]$  in superfluid system play the same role as $[a(\mathbf{x}), b(\mathbf{x})]$ in ``deformed QCD"  model.
  
  In both cases the auxiliary topological fields are effectively static. In case of the ``deformed QCD"  model 
  this is a direct consequence of the construction, see Appendix \ref{deformed}. In case of a superfluid system it is obviously a result of our   assumption formulated after   eq.(\ref{current}). The physical meaning of this assumption is explained at the very end of the last section. The main point is that this approximation is somewhat justified a posteriori by demonstrating that this quasi-static induced background (formulated in terms of   auxiliary fields $[a_k(\mathbf{x}), b_k(\mathbf{x})]$)   indeed generates a finite contribution to $\la I^2\ra_a$.  Therefore, the   auxiliary fields  can be treated as  a  slow effective background which produces a finite contribution to the free energy revealed by the computations of $\la I^2\ra_a$. 
  
  One should comment here that we   refer   in this ``formal similarities" section  to  the weakly coupled gauge theory,  the ``deformed QCD"  model (where analytical computations can be explicitly performed) rather than to strongly coupled QCD. However, it is known  that very similar structure in topological susceptibility $\la Q^2\ra$  also occurs in real strongly coupled QCD, see \cite{Zhitnitsky:2013hs} and references therein. Furthermore, it has been argued \cite{Unsal:2008ch}, that  there is no any phase transition  along the passage from   weakly coupled  ``deformed QCD" model  to    strongly coupled QCD. 
   \subsection{Fundamental difference between QCD and superfluid systems}\label{difference}
These formal similarities, however,   should not hide   a fundamental difference between   superfluid  systems defined in Minkowski space time   and Euclidean 4d 
``deformed QCD"  odel which is reviewed in Appendix \ref{deformed}.
In particular, instead of real vortices in superfluid system we have pseudo-particles (monopoles) which saturate the path integral. 
These pseudo-particles  describe the   time-dependent   tunnelling transitions  between physically identical but topologically distinct  winding $|n\ra$ states connected by the large gauge transformations. These pseudo-particles   
  are not real physical states in Hilbert space, in contrast with vortices in a superfluid system. 
It means that these  pseudo-particles should be treated as a convenient computational technical tool in  evaluation  of  the path integral. It should be  contrasted with  quasi-particles (phonons and rotons) in superfluid systems which  can propagate in physical space-time, and which can be explicitly  observed by studying a  varies correlation functions. 
 
The  observation  of real quasi-particles and real vortices  in a number of different  physical experiments, see textbook \cite{book} for references, should be contrasted with observed topological structure in the path integral computations in QCD using the lattice simulations in 4d Euclidean space-time, see next  section \ref{skeleton} for the details and references. 

However, there is a common denominator between these two very different studies (superfluidity  vs QCD) carried out in two very different space-times 
(Minkowski vs  Euclidean). In both cases there is a  hidden long range order. It is formally expressed in terms of  a non-physical pole at $p^2=0$ in eqs.(\ref{V_a}) and (\ref{A}) in gauge-dependent correlation functions. This pole, being unphysical itself, nevertheless  generates     the physical contact terms  in gauge-invariant  correlation functions $\la I^2\ra$ and $\la Q^2\ra$ correspondingly.  The manifestations of this long range structure is very different for two systems: in superfluid systems this long range structure manifests itself as a presence of a network of long vortices with a typical length scale of order of the  size of a system \cite{vortices}, while in QCD this long range structure can be  observed in lattice simulations where relevant 4D topological configurations are correlated on   scales of order the    the entire lattice \cite{Horvath:2003yj}. We elaborate on this similarity in next subsection \ref{skeleton}.

    \subsection{Network of superfluid  vortices in HeliumII \cite{vortices} vs     ``Skeleton network" in lattice QCD \cite{Horvath:2003yj}}\label{skeleton}
    Before we elaborate on connection between the network of superfluid  vortices (unambiguously  observed in Helium \cite{vortices})
    and a complicated topological structure in QCD (observed in Monte Carlo lattice simulations, dubbed as ``skeleton network"   \cite{Horvath:2003yj}), we want to list the main properties of the ``skeleton network".

 The gauge configurations observed in  \cite{Horvath:2003yj} display a laminar structure in the vacuum consisting of extended, thin, coherent, locally low-dimensional sheets of topological charge embedded in 4d space, with opposite sign sheets interleaved, see original QCD lattice results~\cite{Horvath:2003yj,Horvath:2005rv,Horvath:2005cv,Alexandru:2005bn}.
A similar structure has been also observed in QCD by different groups~\cite{Ilgenfritz:2007xu,Ilgenfritz:2008ia,Bruckmann:2011ve,Kovalenko:2004xm,Buividovich:2011cv}.
Furthermore, the studies of localization properties of Dirac eigenmodes have also shown evidence for the delocalization of low-lying modes on effectively low-dimensional surfaces.
The following is a list of the key properties of these gauge configurations which we wish to review:

1) The tension of the ``low dimensional objects''  vanishes below the critical temperature and these objects percolate through the vacuum, forming a kind of a vacuum condensate; 

2) These ``objects'' do not percolate through the whole 4d volume, but rather, lie on low dimensional surfaces $1\leq d < 4$ which   organize  a coherent   double layer structure; 

3) The total area of the surfaces is dominated by a single percolating cluster of ``low dimensional object''; 

4) The contribution of the percolating objects to the topological susceptibility $\la Q^2\ra$ has the same sign compared to its total value; 

5) The width of the percolating objects apparently vanishes in the continuum limit; 

6) The density of well localized 4d objects (such as small size instantons) apparently vanishes in the continuum limit.
     
    This structure obviously collapses above the phase transition point at $T>T_c$. These drastic  changes at the critical temperature  can be, in fact,  quantitively  understood \cite{Parnachev:2008fy}  in large $N$ limit when the topological susceptibility is  saturated by conventional instantons at $T>T_c$, while it is presumably saturated by      the   ``skeleton network" at $T<T_c$ described above{\footnote{One may wonder how the ``skeleton network" at $T<T_c$ can emerge from  the fractionally charged point-like monopoles of the ``deformed QCD" model? Indeed, it is known that all important features of  the ``deformed QCD" model are saturated by fractionally charged monopoles, see Appendix \ref{deformed} for review. It is also known that the ``deformed QCD" model  becomes the strongly coupled QCD when the size of the compact $\mathbb{S}^1$ adiabatically increases.  The conjecture \cite{Thomas:2012tu} is that some extended  topological objects   which are inevitably present in the ``deformed QCD" model,   eventually form the ``skeleton network" observed in  \cite{Horvath:2003yj,Horvath:2005rv,Horvath:2005cv,Alexandru:2005bn,Ilgenfritz:2007xu,Ilgenfritz:2008ia,Bruckmann:2011ve,Kovalenko:2004xm} when one  slowly moves  from weakly coupled model to strongly coupled QCD by increasing  the size of $\mathbb{S}^1$. The structure  of these objects obviously must drastically change    in passage from weakly coupled to strongly coupled regime. Recent numerical studies \cite{Liu:2015ufa,Liu:2015jsa} of a similar model (where point -like  fractionally charged monopolies saturate the partition function) implicitly support this conjecture. Indeed, the  semiclassical approximation employed in \cite{Liu:2015ufa,Liu:2015jsa} breaks down   when the size of $ \mathbb{S}^1$ increases.  It  can be interpreted as that the elementary point-like monopoles cannot saturate the partition function at sufficiently large size of $ \mathbb{S}^1$ and must transform themselves into a more complicated (extended) objects. The dynamics of this  transformation  (from point-like pseudoparticles to extended structure) is   still not well understood.}. 

It has been argued for quite a long time, see \cite{book} for references,  that the microscopical origin of the phase transition in superfluid systems is  somehow related to the network of vortices. This coherent  network is expected to become very dense, highly knotted, folded and  crumpled when the critical temperature approaches the phase transition point. The superfluid vortices, which represent the main constituent of this network, of course disappear above $T_c$. Entire network is expected to collapse exactly at this point as the building material, the superfluid vortices disappear
at $T=T_c$. 
Details of the dynamics describing this  complicated picture  of the network's evolution  is obviously prerogative of numerical simulations, similar to the QCD studies 
 \cite{Horvath:2003yj,Horvath:2005rv,Horvath:2005cv,Alexandru:2005bn,Ilgenfritz:2007xu,Ilgenfritz:2008ia,Bruckmann:2011ve,Kovalenko:2004xm}. 

 We would like to speculate here  that the structure which has  been observed on the lattice \cite{Horvath:2003yj,Horvath:2005rv,Horvath:2005cv,Alexandru:2005bn,Ilgenfritz:2007xu,Ilgenfritz:2008ia,Bruckmann:2011ve,Kovalenko:2004xm} plays the same role as the network of   vortices experimentally observed  in superfluid He II \cite{vortices}.  In other words, we want to advocate an idea   that the corresponding networks provide the  microscopical  mechanisms responsible  for the phase transitions in both cases: confinement-deconfinement phase transition in the QCD case, and superfluid to the normal liquid phase transition in case of superfluid helium II. 
 
Indeed, in both cases the configurations  themselves  have lower dimensionality than the space itself. However, these low-dimensional  configurations are so dense, and they fluctuate so strongly that they almost fill the entire space. 
In both cases an effective  tension (representing the superposition  of internal tension combined with the entropy) of the configurations vanishes as a result of large entropies of the objects which overcome the internal tension. This leads to the  percolation 
of the  vortices in superfluid He II  and   formation of the the ``Skeleton" in QCD correspondingly. If the effective tensions 
 of these configurations did not vanish, we would observe a finite number of fluctuating objects with finite size in the system instead of observed percolation of the ``Skeleton" and superfluid vortices.
Furthermore,  typically  the ``Skeleton"   spreads  over maximal   distances percolating through the entire volume of the system similar to superfluid vortices.

Our final comment is related to the  ${\cal P}$ invariance in both systems. The question on ${\cal P}$ invariance occurs in QCD due to the fact that the topological density operator $Q$ is a pseudoscalar. A similar  comment also holds for  circulation density field $ \gamma_k(\mathbf{x})$ defined by eq. (\ref{analogy}) which is the axial rather than vector field. Our auxiliary topological field $a_k(\mathbf{x})$ is also axial rather than vector field as the relation (\ref{poisson}) states. 
The correlation functions $\la I^2\ra_a$ is  defined in terms of $a_k(\mathbf{x})$ by eq. (\ref{I_a}), while  $\la Q^2\ra$  in ``deformed QCD" model is defined in terms of $a(\mathbf{x})$ by (\ref{top_QCD}). These correlation functions  are obviously ${\cal P}$-even observables  constructed from ${\cal P}$ -odd objects. 
  In ``Skeleton" studies  \cite{Horvath:2003yj} there are two oppositely- charged sign-coherent connected structures (sheets).  The ${\cal P}$  invariance holds in QCD as a result of  delicate cancellation between the opposite sign interleaved sheets.  
  
  A similar studies on sign of the circulation  in superfluid He II have  not been  performed in \cite{vortices}. However, it is quite obvious that the ${\cal P}$ invariance can  be locally maintained only as a result of similar cancellation between  opposite sign coherent interleaved   vortices with opposite circulations.  
  
  One should remark here that if the external parameter $\theta\neq 0$  is not vanishing  in QCD  than the delicate cancellation mentioned above does not hold anymore, and ${\cal P}$  asymmetry would be  generated in the system\footnote{This is renowned   strong $\cal{CP}$ problem in QCD as the $\theta$ term violates both: ${\cal P}$ and  $\cal{CP}$ symmetries of the theory. The resolution of this problem consists a new fundamental particle, the axion, yet to be discovered,  which dynamically drives  $\theta\rightarrow 0$ during the QCD phase transition in early Universe.}. Similar comment also applies to a superfluid system when the role of $\theta$ plays 
  external parameter $\vec{s}$ defined by eqs. (\ref{s}), (\ref{n_s}). Indeed, up to normalization factors both correlation functions $\la I^2\ra$ and  $\la Q^2\ra$ are expressed as the second derivatives of the partition function ${\cal{Z}}$ with respect to $\vec{s}$ and $\theta$ correspondingly
  \be
  \label{I_Q}
  \la I^2\ra\sim \frac{\partial^2 \ln {\cal{Z}}}{\partial \vec{s}^2}, ~~~ \la Q^2\ra\sim \frac{\partial^2 \ln {\cal{Z}}}{\partial \theta^2}.
  \ee
  It is quite obvious that $ \vec{s}\neq 0$ in superfluid liquid   breaks ${\cal P}$ invariance of  the system (similar to QCD at $\theta\neq 0$) as 
  it corresponds to  a presence of  superfluid flow   in a specific direction according to (\ref{s}).

The crucial difference between the two networks discussed above   is of course the nature of the constituents of  these networks:  superfluid macroscopically large vortices   live in  real Minkowski space-time while lattice QCD measurements are done in Euclidean space-time where the corresponding long ranged configurations    saturate the path integral, and describe the tunnelling processes,  as we already discussed  in section \ref{difference}.     
  
\section{Conclusion. Speculations. }\label{conclusion}
    Our conclusion can be separated into three  related, but still distinct pieces: First,  we highlight the  theoretical results  on computations of $\la I^2\ra$ based on the   path integral approach 
in a superfluid   system. Secondly,  we mention on possible connection of our computations with related works in other fields. 
Finally, we present some  speculations related to strongly coupled QCD   where fundamentally the same    effects do occur, and might be the crucial  ingredients   in understanding of the observed  cosmological  vacuum energy today, the so-called dark energy.

\subsection{ Results.}  The main ``technical" claim  of this work  is that the contribution of   a complicated network of knotted, folded, twisted    and wrinkled vortices to winding number susceptibility $\la I^2\ra$    can be formulated in terms of an auxiliary topological fields 
    $[a_k(\mathbf{x}) , b_k(\mathbf{x})] $. These fields are not new degrees of freedom, and they  do not propagate. Rather, they effectively describe the long-range dynamics of this complicated network of strongly interacting vortices. Though these fields are not independent degrees of freedom, they still produce a number of physical effects. In particular, they generate the contact term in 
    the winding number correlation function $\la I^2\ra$.
   In addition, $a_k(\mathbf{x})$ mixes with the Goldstone field, becomes a massive propagating degree of freedom,   the vorton.    
     
     It is a highly nontrivial phenomenon, and it has its counterpart in relativistic particles physics, where a similar effect does occur, 
     and it represents the resolution of the  celebrated  $U(1)$ problem as formulated by Witten and Veneziano in 1980 \cite{witten, ven}. 
    Precisely the    Veneziano ghost, which plays the role of topological auxiliary fields $[a_k(\mathbf{x}) , b_k(\mathbf{x})] $ in present context,   is the crucial element in the resolution of the $U(1)$ problem and generates the contact term in topological susceptibility  $\la Q^2\ra$ confirmed by numerous lattice simulations.   The Veneziano ghost  does not contribute to any other correlation functions.   We have made a number of comments demonstrating a close 
    relation between  analysis of  $\la I^2\ra$ in superfluid system (discussed  in sections \ref{susceptibility}, \ref{numerics})   and  $\la Q^2\ra$ in a simplified ``deformed QCD" model reviewed  in Appendix \ref{deformed}. 
       
  It might look very suspicious that  a complicated dynamics of network of  the vortices (which is known to emerge in superfluid systems in vicinity of the phase transition) can be described in so simple  way in terms of   local   auxiliary  topological fields $[a_k(\mathbf{x}) , b_k(\mathbf{x})] $. 
  However, such a  complimentary description becomes     less mysterious (but still highly nontrivial) if one  recalls
  that many systems demonstrate the particle-vortex duality  which  has been known  since \cite{Peskin}. Therefore, it is not  really a big surprise that some elements   of an   approximate ``duality" (between superfluid vortices and local topological fields $[a_k(\mathbf{x}) , b_k(\mathbf{x})] $) also emerge in our superfluid system.  
    
  \exclude{ One  should also comment  that an analogous treatment in  solvable ``deformed QCD" model exactly reproduces $\la Q^2\ra$ which were originally derived without even mentioning the auxiliary topological fields.  Furthermore, it is known that exactly the same structure occurs in strongly coupled QCD where it is generated by much more complicated configurations than point-like monopolies of the ``deformed QCD" model. 
 } 
  It is quite obvious that our studies  of $\la I^2\ra$ in terms of  topological auxiliary fields $[a_k(\mathbf{x}) , b_k(\mathbf{x})]$  is only the very first step in direction of complete description of $\la I^2\ra$ which is precisely  related to superfluid density as thermodynamical relation (\ref{n_s}) states. Indeed, we  kept only the quadratic terms in the Hamiltonian by consistently neglecting all the interactions in order to compute  the path integral. Furthermore, we assumed that the winding number correlation function  $\la I^2\ra$  is saturated by
  static  non-propagating configurations $[a_k(\mathbf{x}) , b_k(\mathbf{x})] $,   while the network of vortices is obviously a strongly  fluctuating system. This assumption is somewhat justified a posteriori by demonstrating that such static configurations make a finite contribution to  $\la I^2\ra$. In this sense, this treatment can be interpreted as some kind of  ``mean field" approximation in computing  $\la I^2\ra$.

  Accounting for all these (neglected) interactions is obviously very ambitious task which is well beyond the scope of the present work. However, we expect (based on the experience with computations of  $\la Q^2\ra$  in QCD  and its relation with analogous computations in ``deformed QCD" model) that the corresponding interactions would ``renormalize" the parameters of the system, but not drastically change the structure of   the winding number correlation function (\ref{I_a}), including the contact term,  which was computed  by keeping the dominant quadratic terms and neglecting all sub-leading (at large distances)    interacting  terms\footnote{It is quite possible that   some previously developed models such as ``Vortex-Ring Model" from                   refs.\cite{Williams,Shenoy}, might be useful in making a next step in this direction.}. 
  
  \subsection{\label{relation}Relations to other approaches.} We want to make few additional comments on possible relation of our treatment of   $\la Q^2\ra$ with other computations  where similar structure 
   for topological susceptibility $\la Q^2\ra$ is known to occur. First of all we have in mind the saturation of the topological susceptibility $\la Q^2\ra$  by the Veneziano ghost \cite{ven} which can be understood in terms of auxiliary topological fields 
    $[a(\mathbf{x}) , b(\mathbf{x})] $ in our computations as reviewed  in Appendix \ref{deformed}. The key observation relevant for the present work is that the corresponding contact term must  be related somehow  to the  so-called  Gribov's copies \cite{Gribov}.   Indeed, in the Coulomb gauge the emergence of the Gribov's copies can be traced to the presence of the topological sectors in  a gauge theory. The tunnelling transitions between these topological sectors saturate the contact term in topological susceptibility  as reviewed  in Appendix \ref{deformed}. Therefore, the contact term, which is the key element in our analysis (based on formulation  in terms of the auxiliary topological fields)    must be  related to the Gribov's copies which describe this type of  physics. 
    
    Furthermore, it has been recently argued that the Gribov's copies are also  inherently related to  confinement in   QCD, see \cite{Kharzeev:2015xsa} and references on related works therein. Such a relation strongly suggests that the auxiliary topological fields (which is the main technical tool in our approach) are ultimately related to the nature of the confinement of the theory as they saturate the topological susceptibility  $\la Q^2\ra$. 
    
    In the ``deformed QCD" model 
  reviewed  in Appendix \ref{deformed}  all these relations  and connections  is almost a trivial remark as  the topological auxiliary fields effectively describe the dynamics of the monopoles which indeed provide the confinement in the theory. However, our claim is much  more generic and applies to the strongly coupled gauge theory where explicit relation between the two descriptions is far from obvious. The phase transition formulated in  terms of the local auxiliary  fields   implies that these topological fields experience the  drastic modifications    in   vicinity of the phase transition. 
  
  In context of the present work where the phase transition in He-II can be understood in terms of the  percolating network  of vortices  the corresponding  drastic changes indeed  take place in the system as we discussed in section \ref{numerics}. If we assume that the analogy between superfluid He II  and the QCD vacuum is sufficiently deep,  as argued in section \ref{interpretation}  then 
  one may ask the following question:    what kind of objects play the role of superfluid vortices  in the QCD phase transition?
  One can rephrase the same question in slightly different way: it is known that in ``deformed QCD" model the corresponding 
  correlation function $\la Q^2\ra$ and the string tension are saturated by the monopoles. What happens to these abelian monopoles when one slowly moves     from weakly coupled ``deformed QCD" to strongly coupled QCD? 
 
  We obviously do not know a precise  answer to this question. However, we strongly suspect than the centre vortices, see   \cite{Greensite:2011zz}  for review,  which have been  observed in numerous  lattice simulations (and which generate  the dominant contribution   to the string tension and topological susceptibility in the confined phase)  may  replace the monopoles which are dominant pseudoparticles in ``deformed QCD" model.   In fact there is some numerical analysis   \cite{deForcrand:2000pg} suggesting that 
  indeed there is a close connection between these two descriptions (in terms of monopoles and centre vortices) when monopoles and anti-monopoles look like alternating beads on a necklace formed by the centre vortex ring. 
  
 We conclude this subsection with the following comment. The dual description in both cases (superfluid system  and QCD) is formulated in terms of local auxiliary topological fields:  $[a(\mathbf{x}) , b(\mathbf{x})] $ in  ``deformed QCD" (and corresponding generalization in terms of the Veneziano ghost in  strongly coupled QCD), and   $[a_k(\mathbf{x}) , b_k(\mathbf{x})] $   in superfluid system. The corresponding similarities and analogies  between the two systems have been extensively discussed in section \ref{susceptibility}. The descriptions of the two systems in terms of the original variables (superfluid vortices in He-II and, possible  centre vortices in QCD) most likely are very different;  it is just the dual descriptions formulated in terms of the auxiliary topological fields in two drastically different systems look very much the same. 
  
We do not know how  the similarities observed in  the dual descriptions in sections \ref{susceptibility}, \ref{numerics},  \ref{interpretation} may manifest themselves  if  the systems are formulated  in original terms rather than in dual variables.  
  Nevertheless, the IR  pole in  (\ref{V_a}) and (\ref{A}) in gauge-dependent correlation functions (in superfluid system and QCD correspondingly) and related  contact terms in gauge invariant correlation functions  must be somehow generated
when computations are done in terms of the original variables  as well.

\subsection{Speculations.} The final part of our conclusion is much more  speculative, but we think it is worthwhile to mention these  speculations because  they  may have profound consequences on  our understanding of the nature of the vacuum energy of the Universe we live in. 
       
       The key point of our analysis is  the presence of the IR pole at $p^2=0$ in gauge dependent correlation functions in superfluid system (\ref{V_a}) and analogous relation  (\ref{A}) in QCD. While there are no any physical massless degrees of freedom associated with this pole,   
       it still generates a physical contribution to gauge invariant correlation functions in form of the contact terms  (\ref{I_a}) and  (\ref{QCD_final}) in superfluid  liquid   in QCD correspondingly. The crucial  point is that the corresponding contact term is proportional to $  \delta(\bold{x})$ function. However, it   is generated by IR rather than UV physics, which could be naively identified  with  $  \delta(\bold{x})$-like behaviour.
       In other words, the contact term should be interpreted as   a surface integral which is highly sensitive to the long- distances IR physics, i.e. 
       \be
     \int d^3x   \delta(\bold{x})\sim \int d^3x ~\partial_{i}\left(\frac{x_i}{4\pi x^3}\right)\sim \oint_S dS^{i}\left(\frac{x_i}{4\pi x^3}\right).~~
       \ee
     This interpretation fits nicely  with  microscopical picture we are advocating in this work  that the corresponding contact term is generated by a long -ranged network of macroscopically large  vortices in superfluid system and by the ``Skeleton" in QCD as we discussed in  \ref{skeleton}. 
     
     If we accept this interpretation,  suggesting the IR nature of the contact term (and   the energy (\ref{I_Q}) associated with the contact term), one should also accept a direct consequence of such interpretation that the  energy density of the system  in the bulk might be highly sensitive to the boundary conditions, even  though  there are no any massless physical propagating degrees of freedom responsible for  such sensitivity, in  very much the same way as it happens in topological insulators.  We  reiterate this 
     statement as follows: there is an extra energy  in the bulk  of the   system associated with the contact terms, which however, cannot be expressed in terms of any physical propagating degrees of freedom. 
            
In   QCD context the presence  of   the  vacuum energy  not related to   any physical propagating degrees of freedom was the main  motivation for the proposal   \cite{Zhitnitsky:2013pna,Zhitnitsky:2015dia}  that the  observed dark  energy in  the Universe may have, in fact,  precisely such non-dispersive  nature\footnote{This novel type  of vacuum energy which can not be expressed in terms of propagating degrees of freedom has in fact been well studied in QCD lattice simulations, see \cite{Zhitnitsky:2013pna} with a large number of references on the original lattice results.}. This proposal where an extra energy  cannot be associated with any propagating particles  should be contrasted with a commonly accepted paradigm   when an extra vacuum energy in the Universe is always associated with some  ad hoc    propagating degree of freedom\footnote{There are two instances in the evolution of the Universe when the vacuum energy plays a crucial  role.
The first instance   is identified with  the inflationary epoch  when the Hubble constant $H$ was almost constant, which corresponds to the de Sitter type behaviour $a(t)\sim \exp(Ht)$ with exponential growth of the size $a(t)$ of the Universe. The  second instance where the vacuum energy plays a dominant role  corresponds to the present epoch when the vacuum energy is identified with the so-called dark energy $\rho_{DE}$ which constitutes almost $70\%$ of the critical density. In the proposal  \cite{Zhitnitsky:2013pna,Zhitnitsky:2015dia}  the vacuum energy density can be estimated as $\rho_{DE}\sim H\Lambda^3_{QCD}\sim (10^{-4}{\rm  eV})^4$, which is amazingly  close to the observed value.}.

   In the superfluid context the presence of the contact term  which cannot be associated with any propagating degrees of freedom implies that 
   some observables such as superfluid density $n_s$ are algebraically (rather than exponentially) sensitive to the boundary conditions. In other words, it is naturally to expect that  the superfluid density $n_s (L)$ in a container with typical size $L$ is sightly different from $n_s (L=\infty)$, i.e.
   \be
   \label{L}
   n_s (L)= n_s (L=\infty)\left[1+{\cal{O}}\left(\frac{1}{L}\right)\right], 
   \ee 
  similar to computations in ``deformed QCD" model \cite{Thomas:2012ib}. 
     It could be interpreted as a Topological Casimir Effect (TCE)  when nontrivial topology may result in additional energy in the bulk of the  system which cannot be expressed in terms of conventional propagating degrees of freedom, but rather is sensitive to topological features of the system \cite{Cao:2013na,Yao:2016bps}. 
     
     This effect (\ref{L}) in many respects is very similar to TCE in  the Maxwell theory   formulated on a compact manifold with nontrivial  $\pi_1[U(1)]=\mathbb{Z}$ when the extra term for vacuum energy is generated due to the presence of the winding states $|n\ra$ and tunnelling transitions between them, see   \cite{Cao:2013na,Yao:2016bps} for the details. The nature of this extra vacuum energy is very different  
  from conventional Casimir Effect which is generated  due to the conventional fluctuations of the physical photons with two transverse  polarizations between two neutral plates with trivial topology. 
  
  Furthermore, this extra energy can be also formulated \cite{Yao:2016bps} in terms of auxiliary topological fields which cannot be  expressed in terms of the  physical propagating photons with two transverse polarizations. These topological fields are,  in fact,  analogous to auxiliary fields $[a_k(\mathbf{x}) , b_k(\mathbf{x})]$  introduced in the present work. More than that, the  classification of the topological objects from refs. \cite{Cao:2013na,Yao:2016bps}
  is based on $\pi_1[U(1)]=\mathbb{Z}$    which is equivalent to  classification of  the vortices in superfluid systems which is the main subject of  the present work. 
     
     Essentially, eq. (\ref{L})  suggests that one can use a superfluid system  to study very deep and intriguing features of the QCD vacuum, in spite of the huge differences in nature of these systems as discussed in Section \ref{difference}. In particular, the relevant structure in a superfluid  system is a complicated net of vortices in Minkowski space-time, while in QCD it is a similar complicated net of configurations describing the tunnelling transitions in Euclidean space-time. Nevertheless, the key element in  both systems, as emphasized in section \ref{skeleton},  is there existence of a net which eventually generates    the contact term  (and   the  long-range sensitivity of some  observables
     related to this contact term)   in the bulk of the system.
  
     The corresponding contributions   to the energy in many systems (such as superfluid liquid (\ref{L}), QCD \cite{Zhitnitsky:2015dia},  ``deformed QCD" model \cite{Thomas:2012ib},  Maxwell  theory on a compact manifold \cite{Cao:2013na}) are fundamentally not expressible   in terms of any physical propagating degrees of freedom. Rather, these terms  reflect the    topological features of a system. 
     
   $\bullet$    To conclude:  In this work we presented one more example which shows amazing conceptual similarity between particle physics and a superfluid system. Our  hope is that this work  may generate future studies    benefiting  both fields.  

   \section*{Acknowledgments}
 I am very thankful to Sasha Gorsky and Dima Kharzeev  for the invaluable discussions  (which essentially initiated this project) during a workshop 
 at the Simons Center for geometry and physics in May 2015. I am also thankful to  Philippe  de Forcrand,  Jeff Greensite  and Tin Sulejmanpasic and other participants of the workshop ``gauge topology: from lattice to collider", ECT$^*$, Trento, November 2016,
where this work has been presented,   for   useful   discussions a  comments. 
This work was supported in part by the National Science and Engineering
Research Council of Canada. 

\appendix
\section{Topological susceptibility in ``deformed QCD".  The lessons for superfluid systems.}\label{deformed}
The main goal of this Appendix is to demonstrate a close analogy between  the computations of $\la I^2\ra$ for superfluid system presented in the main text and computations of $\la Q^2\ra$ in QCD. Furthermore, we also show that  the mass scale (\ref{gap}) which is generated in the superfluid system is very similar to the mass scale $m_{\eta'}$ which is generated in QCD. In both cases the new mass scales manifest themselves in  studying very specific correlation functions: $\la I^2\ra$  in superfluid systems,  and  $\la Q^2\ra$ in QCD.

To proceed with this task  we compute in  this Appendix the topological susceptibility $\la Q^2\ra$ in ``deformed QCD" model developed in \cite{Unsal:2008ch}. 
 This   is a weakly coupled gauge theory, but nevertheless  preserves all the crucial elements of strongly interacting QCD, including confinement, nontrivial $\theta$ dependence, degeneracy of the topological sectors,  etc.  The topological susceptibility $\la Q^2\ra$ plays a key role in resolution of the celebrated  $U(1)_A$  problem \cite{witten, ven, vendiv} where the would be $\eta'$ Goldstone boson generates its mass   as a result of mixing of the Goldstone field with a topological auxiliary field characterizing the system. 
 
 One should emphasize that all the elements of this mechanism  are well supported by the lattice simulations in strongly coupled QCD. However, in this Appendix we use  the weakly coupled ``deformed QCD" model to demonstrate how all the crucial elements work in this mechanism using analytical computations.

The plan of this Appendix is as follows. The basics of this model are reviewed in section \ref{model}.
 In section \ref{Appendix:mass}   we explain  how the  Goldstone boson generates its mass   as a result of mixing of the Goldstone field with a topological auxiliary field. Finally, in section  \ref{ghost} we reformulate our results to make very close connection with Veneziano ghost description. From this complimentary analysis one should be clear that the topological auxiliary fields are not present in the Hilbert space as the asymptotic states, and these auxiliary fields  do not violate any fundamental principles of the theory.

\subsection{The Model}\label{model}
In the deformed theory an extra ``deformation'' term is put into the Lagrangian as suggested in  \cite{Unsal:2008ch} in order to prevent the center symmetry breaking that characterizes the QCD phase transition between ``confined'' hadronic matter and ``deconfined'' quark-gluon plasma, thereby explicitly preventing that transition.
We start with pure Yang-Mills (gluodynamics) with gauge group $SU(N)$ on the manifold $\mathbb{R}^{3} \times S^{1}$ with the standard action
\be \label{standardYM}
	S^{YM} = \int_{\mathbb{R}^{3} \times S^{1}} d^{4}x\; \frac{1}{2 g^2} \mathrm{tr} \left[ F_{\mu\nu}^{2} (x) \right],
\ee
and add to it a deformation action,
\be \label{deformation}
	\Delta S \equiv \int_{\mathbb{R}^{3}}d^{3}x \; \frac{1}{L^{3}} P \left[ \Omega(\mathbf{x}) \right],
\ee 
built out of the Wilson loop (Polyakov loop) wrapping the compact dimension
\be \label{loop}
	\Omega(\mathbf{x}) \equiv \mathcal{P} \left[ e^{i \oint dx_{4} \; A_{4} (\mathbf{x},x_{4})} \right].
\ee
The parameter  $L$ here is the length of the compactified dimension which is assumed to be small. 
The coefficients of the polynomial $P \left[ \Omega(\mathbf{x}) \right]$ can be suitably chosen such that the deformation potential (\ref{deformation}) forces unbroken symmetry at any compactification scales.
At small compactification $L$ the gauge coupling is small so that the semiclassical computations are under complete theoretical control \cite{Unsal:2008ch}.
The proper infrared description of the theory is a dilute gas of $N$ types of monopoles, characterized by their magnetic charges, which are proportional to the simple roots and affine root $\alpha_{a} \in \Delta_{\mathrm{aff}}$ of the Lie algebra for the gauge group $U(1)^{N}$.
For a fundamental monopole with magnetic charge $\alpha_{a} \in \Delta_{\mathrm{aff}}$ (the affine root system), the topological charge is given by
\be \label{topologicalcharge}
	Q = \int_{\mathbb{R}^{3} \times S^{1}} d^{4}x \; \frac{1}{16 \pi^{2}} \mathrm{tr} \left[ F_{\mu\nu} \tilde{F}^{\mu\nu} \right]
		= \pm\frac{1}{N},
\ee
and the Yang-Mills action is given by
\be \label{YMaction}
	S_{YM} = \int_{\mathbb{R}^{3} \times S^{1}} d^{4}x \; \frac{1}{2 g^{2}} \mathrm{tr} \left[ F_{\mu\nu}^{2} \right] = \frac{8 \pi^{2}}{g^{2}} \left| Q \right|.
\ee
The $\theta$-parameter in the Yang-Mills action can be included in conventional way,
\be \label{thetaincluded}
	S_{\mathrm{YM}} \rightarrow S_{\mathrm{YM}} + i \theta \int_{\mathbb{R}^{3} \times S^{1}} d^{4}x\frac{1}{16 \pi^{2}} \mathrm{tr}
		\left[ F_{\mu\nu} \tilde{F}^{\mu\nu} \right],
\ee
with $\tilde{F}^{\mu\nu} \equiv \epsilon^{\mu\nu\rho\sigma} F_{\rho\sigma}$.

The system of interacting monopoles, including the $\theta$ parameter, can be represented in the  sine-Gordon form as follows \cite{Unsal:2008ch,Thomas:2011ee},
\be
\label{thetaaction}
	S_{\mathrm{dual}} &=& \int_{\mathbb{R}^{3}}  d^{3}x \frac{1}{2 L} \left( \frac{g}{2 \pi} \right)^{2}
		\left( \nabla \bm{\sigma} \right)^{2} \nonumber \\
	    &-& \zeta  \int_{\mathbb{R}^{3}}  d^{3}x \sum_{a = 1}^{N} \cos \left( \alpha_{a} \cdot \bm{\sigma}
		+ \frac{\theta}{N} \right), 
\ee
where $\zeta$ is magnetic monopole fugacity which can be explicitly computed in this model using the conventional semiclassical approximation.
The $\theta$ parameter enters the effective Lagrangian (\ref{thetaaction}) as $\theta/N$ which is the direct consequence of the fractional topological charges of the monopoles (\ref{topologicalcharge}).
Nevertheless, the theory is still $2\pi$ periodic.
This $2\pi$ periodicity of the theory is restored not due to the $2\pi$ periodicity of Lagrangian (\ref{thetaaction}).
Rather, it is restored as a result of summation over all branches of the theory when the levels cross at $\theta=\pi (mod ~2\pi)$ and one branch replaces another and becomes the lowest energy state as discussed in \cite{Thomas:2011ee}.
 
Finally, the dimensional parameter which governs the dynamics of the problem is the Debye correlation length of the monopole's gas, 
\be \label{sigmamass}
	m_{\sigma}^{2} \equiv L \zeta \left( \frac{4\pi}{g} \right)^{2}.
\ee
The average number of monopoles in a ``Debye volume'' is given by
\begin{equation} \label{debye}
	{\cal{N}}\equiv	m_{\sigma}^{-3} \zeta = \left( \frac{g}{4\pi} \right)^{3} \frac{1}{\sqrt{L^3 \zeta}} \gg 1,
\end{equation} 
The last inequality holds since the monopole fugacity is exponentially suppressed, $\zeta \sim e^{-1/g^2}$, and in fact we can view (\ref{debye}) as a constraint on the region validity where semiclassical approximation is justified. This parameter ${\cal N}$ is therefore one measure of ``semi-classicality''.

\subsection{How the Goldstone field receives its mass}\label{Appendix:mass}
   In the  sine-Gordon formulation  (\ref{thetaaction})  the $\eta'$ meson field   appears  exclusively in combination with the $\theta$ parameter as $\theta \rightarrow \theta - \phi (x)$, where $\phi$ is the phase of the chiral  condensate which, up to dimensional normalization parameter,  is identified with physical  $\eta'$ meson in QCD. As it is well known, this property is the direct result of the transformation properties of the path integral measure under the chiral transformations $\psi\rightarrow \exp(i \gamma_5\frac{\phi}{2})\psi$.  Therefore, $\phi (x) $ enters the effective action  (\ref{thetaaction})  exactly in the combination  $\left(\theta- \phi(\mathbf{x})\right)/N$ when we include light quarks into the system.
   
   The next step in our presentation is the computation of the topological susceptibility in this model.
   These  computations can be explicitly carried out as the system is weakly coupled gauge theory, and the semiclassical approximation under complete theoretical control.
   The result of this  computation is    \cite{Thomas:2011ee}:
   \be
\label{QCD_top}
\la q(\mathbf{x}), q(\mathbf{0})\ra_{QCD} =\frac{\zeta}{NL^2}\left[ \delta(\bold{x})-m_{\eta'}^2\frac{e^{-m_{\eta'}r}}{4\pi r} \right],~~
\ee
where new mass scale $m_{\eta'}$  is generated in the problem and is determined in terms of the original dimensional parameters  $\zeta$ and $L$, 
\be
\label{eta'}
m_{\eta'}^2=\frac{\zeta}{cN}, ~~~~~~ \frac{c}{L}=f_{\eta'}^2
\ee

Originally, formula (\ref{QCD_top}) was derived in \cite{Thomas:2011ee} by explicit integration over all possible monopole's configurations, summing over all possible their orientations in the gauge group $SU(N)$, etc. 
However, we want to derive this formula using a different technique in terms of the auxiliary topological fields, similar to the  
$[a_k(\mathbf{x}), b_k(\mathbf{x})]$ introduced in  section \ref{auxiliary}. Such a computation would  demonstrate  a close  connection  with the analysis  of the winding number correlation function  $\la I^2\ra$  computed in section \ref{winding}.

In fact, the corresponding computations have been already carried out in \cite{Zhitnitsky:2013hs}, and we present  the relevant formulae  in this Appendix. The basic technical idea is exactly as presented in sections   \ref{auxiliary}, \ref{winding} when the auxiliary topological fields (treated  as the slow degrees of freedom)  are introduced into the system by inserting the functional $\delta$ function similar to eq. (\ref{delta}). The next step is 
 to  integrate out the fast degrees of freedom in the background of slow auxiliary topological fields.
 This model is a weakly coupled gauge theory. Therefore, these computations can be carried out exactly. 
 The result is  \cite{Zhitnitsky:2013hs}:
\be
\label{eta-action}
&{\cal Z}&\sim\int {\cal{D}}[b]{\cal{D}}[\bm{\sigma}]{\cal{D}}[ a]{\cal{D}}[\phi]e^{-\left(S_{\rm top}+S_{\rm dual}[\bm{\sigma}, b, \phi]+S_{\phi}\right)}~~~\\
&S&_{\phi}= \int_{\mathbb{R}^{3}}  d^{3}x \frac{c}{2}\left( \nabla \phi \right)^{2}\nonumber\\
&S&_{\rm top}[b, a]=
 \frac{-i }{4 \pi N}  \int_{\mathbb{R}^{3}}  d^{3}x     b(\mathbf{x})\vec{\nabla}^2 a (\mathbf{x}) ;
\nonumber \\
	&S&_{\rm dual}[\bm{\sigma},   b, \phi] = \int_{\mathbb{R}^{3}}  d^{3}x \frac{1}{2 L} \left( \frac{g}{2 \pi} \right)^{2}
		\left( \nabla \bm{\sigma} \right)^{2} \nonumber \\&-& \zeta  \int_{\mathbb{R}^{3}}  d^{3}x \sum_{a = 1}^{N} \cos \left( \alpha_{a} \cdot \bm{\sigma}
		+ \frac{\theta+b(\mathbf{x})-\phi(\mathbf{x}) }{N} \right),
		 \nonumber
\ee
  This formula is analogous to expression (\ref{Z-final}) derived for superfluid system. However, there is a fundamental difference between these two computations. Formula (\ref{eta-action}) is an exact result of computations in semiclassical approximation in a weakly coupled gauge theory. Therefore, all terms, including the interaction terms  are accounted in  (\ref{eta-action}). 
  
  It should be contrasted with formula (\ref{Z-final}) derived for superfluid system. In that case we did  not sum over all  possible  configurations, including all  geometries and topologies of a complicated     network of knotted, folded and wrinkled vortices. Instead, we used a gauge invariance arguments to restore the structure of the quadratic portion of the Hamiltonian 
  (\ref{Z-final}). We neglected all the interactions in partition  function (\ref{Z-final}), see few comments on this simplifications at the end of section \ref{winding}. Another difference between (\ref{eta-action}) and  (\ref{Z-final}) is that the auxiliary fields $[b(\mathbf{x}), a(\mathbf{x})]$  in the deformed QCD model are scalars, while in superfluid system the auxiliary fields  $[b_k(\mathbf{x}), a_k(\mathbf{x})]$ are vectors. The difference can be traced from the fact that the relevant objects in the deformed QCD model  are point-like monopoles, while in superfluid systems the relevant objects are the vortices which are classified by a specific direction of the circulation field $\gamma_k(\mathbf{x})$.

We now return to analysis of the deformed QCD model.  
Our task now is to compute the topological susceptibility using the partition function  (\ref{eta-action}).
The effective action assumes the following form
 \be
 \label{top_QCD}
&\,&\la q(\mathbf{x}), q(\mathbf{0})\ra_{QCD}=\frac{1}{{\cal Z}}  \int \frac{{\cal{D}}[ a]e^{-S_{QCD}}\left[\vec{\nabla}^2 a (\mathbf{x}), \vec{\nabla}^2 a (\mathbf{0})\right] }{\left(4 \pi NL\right)^2} \nonumber\\
&\,&S_{QCD}[a, \phi]=   \frac{1}{2N\zeta (4\pi)^2} \int_{\mathbb{R}^{3}}  d^{3}x  \left[a (\mathbf{x})\vec{\nabla}^2\vec{\nabla}^2 a (\mathbf{x})\right] \nonumber\\
&\,&+ \int_{\mathbb{R}^{3}}  d^{3}x \left[\frac{c}{2}\left( \vec{\nabla} \phi(\mathbf{x}) \right)^{2}+\frac{i}{4\pi N} \vec{\nabla}\phi (\mathbf{x})\cdot \vec{\nabla} a (\mathbf{x})\right]   ,
  \ee
 which should be compared with similar expressions (\ref{I5}), (\ref{H}) derived for superfluid system.  
 
The next step in computation of the topological susceptibility  is the diagonalization of the action to eliminate the non-diagonal  term $ \int  d^{3}x \vec{\nabla}\phi \cdot \vec{\nabla} a$ in (\ref{top_QCD}) by making a shift 
\be
\label{phi_2}
\frac{\phi_2 (\mathbf{x})}{\sqrt{c}} \equiv \phi (\mathbf{x})+\frac{i}{4\pi c N}a (\mathbf{x}).
\ee
 The problem  is reduced to the computations of the Gaussian integral  with  the effective action (after the rescaling),  taking the form
  \be
   \label{S_QCD}
S_{QCD}[a' , \phi_2] &=&  \frac{1}{2} \int_{\mathbb{R}^{3}}  d^{3}x\left( \vec{\nabla} \phi_2(\mathbf{x}) \right)^{2}\\
&+&\frac{1}{2 }  \int_{\mathbb{R}^{3}}  d^{3}x a' (\mathbf{x})\left[ \vec{\nabla}^2\vec{\nabla}^2- m_{\eta'}^2 \vec{\nabla}^2 \right]a' (\mathbf{x})\nonumber
    \ee
 which plays the same role as the Hamiltonian (\ref{4-derivatives}) in our computations for superfluid system. In formula (\ref{S_QCD}) parameter $m_{\eta'}^2$ is the $\eta'$ mass in this model and it is  given by eq.(\ref{eta'}). In terms of this rescaled field  $a' (\mathbf{x})$   the  Gaussian integral which enters (\ref{top_QCD}) can be easily computed and it is given by
\be
\label{top_4}
  \frac{\int{\cal{D}}[ a']e^{-S_{QCD}}\vec{\nabla}^2 a' (\mathbf{x}), \vec{\nabla}^2 a' (\mathbf{0})}{\int{\cal{D}}[ a']e^{-S_{QCD}[a']}}= \left[ \delta(\bold{x})-m_{\eta'}^2G_{m_{\eta'}}(\bold{x}) \right]   \nonumber
\ee
where $S_{QCD}$ is defined by (\ref{S_QCD}) and  the massive Green's function
 $G_{m_{\eta'}}(\bold{x})=\frac{e^{-m_{\eta'}r}}{4\pi r} $ is normalized in conventional way ($m_{\eta'}^2\int d^3xG_{m_{\eta'}}(\bold{x})=1$). Collecting all numerical coefficients from  (\ref{top_QCD})  and (\ref{top_4}) the final expression for the topological susceptibility in the presence of massless quark    takes the form 
 \be
\label{QCD_final}
\la q(\mathbf{x}), q(\mathbf{0})\ra_{QCD} =\frac{\zeta}{NL^2}\left[ \delta(\bold{x})-m_{\eta'}^2\frac{e^{-m_{\eta'}r}}{4\pi r} \right]. ~~~~
\ee
This precisely reproduces our previous formula (\ref{QCD_top}) which was derived by explicit integration over all possible monopole's configurations
without even mentioning the topological auxiliary fields.  The celebrated  $U(1)_A$ problem
is resolved in this framework exclusively as a result of dynamics of the topological  $a(\mathbf{x}), b(\mathbf{x})$ fields. These fields are not propagating degrees of freedom, but nevertheless  
generate  a crucial non-dispersive contribution with the ``wrong sign" which is the key element for the formulation and resolution of the $U(1)_A$ problem and the generation of the $\eta'$ mass. Formula (\ref{QCD_final}) has precisely the same structure as eq. (\ref{I_a}) in our studies on superfluidity. 

The main lesson for our present studies of the superfluid system developed in sections \ref{auxiliary}, \ref{winding}, \ref{mass}  is as  follows. The key correlation function (\ref{QCD_top}), (\ref{QCD_final}) can be computed using two different techniques. First, one can use an explicit direct computation which requires   an explicit summation and integration  over all possible monopole's configurations, including positions and orientations,  to arrive to   (\ref{QCD_top}). This is straightforward  but technically very involved procedure. The second technique uses the auxiliary topological fields which effectively account for 
  the long range dynamics of the monopole's configurations. The corresponding formula (\ref{QCD_final}) exactly reproduces the direct computations (\ref{QCD_top}). The lesson is: this  procedure with auxiliary fields  serves   as a test and   gives us a confidence that this formal  manipulations with the auxiliary fields  reproduces the correct results. 

$\bullet$In our present studies  of superfluid system we do not have a proper technique   capable  for direct computations of the winding number susceptibility $\la I^2\ra$ by summing over all possible 
vortex configurations including all  geometries and topologies of a complicated     network of knotted, folded and wrinkled vortices. Therefore, we used   in sections \ref{auxiliary}, \ref{winding}  a second technique which includes the auxiliary topological fields. We tested this technique in the ``deformed QCD" model. Therefore, we are quite confident in qualitative picture developed in sections \ref{auxiliary}, \ref{winding}, including the generation of the new mass scale (\ref{4-derivatives}), (\ref{gap}) and the contact term (\ref{I_a}).

\subsection{ Topological fields and  the Veneziano ghost.}\label{ghost}  
The main goal of this subsection is to argue that unphysical pole (\ref{V_a}) which was found in our studies in section \ref{mass} is absolutely harmless and it does not correspond to any unphysical or ghost-like behaviour in the system.   We use the ``deformed QCD" model (where all computations can be carried out exactly) to support our claim.  In addition, 
the expression for the correlation  function (\ref{top_4}) with action (\ref{S_QCD}) can be represented in a complementary  way which makes the
connection between the Veneziano ghost and topological fields much more explicit and precise. 
 
 To proceed with out task,  we use a standard  trick to represent the 4-th order operator $\left[ \vec{\nabla}^2\vec{\nabla}^2- m_{\eta'}^2 \vec{\nabla}^2 \right]$ which enters the effective action (\ref{S_QCD}) 
    as a combination of  two terms with the opposite signs: a ghost field $\phi_1$ 
and a massive physical $\hat\phi $ field. To be more specific, we write 
\be
\label{inverse}
\frac{1}{\left[ \vec{\nabla}^2\vec{\nabla}^2- m_{\eta'}^2 \vec{\nabla}^2 \right]}= \frac{1}{m_{\eta'}^2}\left(\frac{1}{ \vec{\nabla}^2 - m_{\eta'}^2 } -  \frac{1}{ \vec{\nabla}^2 } \right), ~~
\ee
such that the Green's function for the $a(\mathbf{x})$ field which enters the expression for the topological susceptibility (\ref{top_4}) can be represented as a combination of two Green's functions,  for the physical massive field with conventional kinetic term and for the ghost field with the ``wrong" sign for the kinetic term. 
This formula is identically the same as (\ref{inverse1}) which was employed in the main text on superfluidity.

 Naively, the presence of 4-th order operator in eq. (\ref{S_QCD})  is a signal that the ghost is present in the system. This signal   is explicit  in eq. (\ref{inverse}).    The   contact term     in this framework is  represented 
by the ghost contribution.  
 Indeed,   the   relevant correlation function   which enters the expression for the topological susceptibility (\ref{top_4}) can be explicitly computed using expression (\ref{inverse}) for the  inverse operator  as follows 
 \be\label{QCD_ghost}
&\,& \frac{\int{\cal{D}}[ a']e^{-S_{QCD}[a']}\vec{\nabla}^2 a' (\mathbf{x}), \vec{\nabla}^2 a' (\mathbf{0})}{\int{\cal{D}}[ a']e^{-S_{QCD}[a']}} \\
&=& \int \frac{\dd^3p  }{\left(2\pi\right)^3} e^{-i p x} \frac{p^4}{m_{\eta'}^2}\left[ - \frac{1}{p^2+m_{\eta'}^2} + \frac{1}{p^2} \right] 
\nonumber \\
   &=&\int \frac{\dd^3p}{\left(2\pi\right)^3}  e^{-i p x} \left[ \frac{p^2}{p^2+m_{\eta'}^2}   \right] = \left[ \delta(\bold{x})-m_{\eta'}^2\frac{e^{-m_{\eta'}r}}{4\pi r} \right],  \nonumber
  \ee
 which, of course, is the same final expression  we had before (\ref{top_4}) with the only difference being that it is now explicitly expressed  as a combination of two terms: a physical massive $\eta'$ contribution and an unphysical contribution which saturates the contact term with the ``wrong" sign.
 
 Such interpretation can be supported by computing $\la a' (\mathbf{x}),  a' (\mathbf{0})\ra$ itself, similar to our studies of the correlation function (\ref{V_a}). 
 \be
 \label{A}
&& \la a' (\mathbf{x}),  a' (\mathbf{0})\ra=
  \frac{\int{\cal{D}}[ a']e^{-S_{QCD}[a']}  a' (\mathbf{x}),   a' (\mathbf{0})}{\int{\cal{D}}[ a']e^{-S_{QCD}[a']}} \nonumber \\
&=& \int \frac{\dd^3p  }{\left(2\pi\right)^3} e^{-i p x} \frac{1}{m_{\eta'}^2}\left[ - \frac{1}{p^2+m_{\eta'}^2} + \frac{1}{p^2} \right] 
     \ee   
 Naively, the presence of the pole at $p^2=0$ with a ``wrong sign" in eq. (\ref{A}) is a signal of a ghost in the system, which implies the violation of  unitarity along with  other fundamental principles of quantum field theory (QFT). 
 Nevertheless, we know that the original theory  is perfectly defined QFT, and the generation of this unphysical pole is simply an artifact of our formal procedure, when we inserted auxiliary topological fields into the system.
 Indeed, we observed above that one computes a gauge invariant correlation function (\ref{QCD_ghost})
 this ``wrong sign" contribution generates the contact term, and it does not correspond to any propagation of any physical degrees of freedom. Another way to support this claim is to construct the physical Hilbert space for the problem, which is our next exercise.  
 
 To proceed with this task  we  represent the  correlation function  (\ref{QCD_ghost}) by introducing two fields $\phi_1(\mathbf{x})$ and $\hat{\phi}(\mathbf{x})$
 replacing the $a'(\mathbf{x})$ which enters the effective action (\ref{S_QCD}) as the 4-th order operator.  To be more precise, we rewrite our action (\ref{S_QCD}) in terms of these new fields  $\phi_1(\mathbf{x})$ and $\hat{\phi}(\mathbf{x})$
 as follows
 \be
   \label{S_ghost}
&&S_{QCD}[\hat{\phi}, \phi_1,\phi_2] =  \frac{1}{2} \int_{\mathbb{R}^{3}}  d^{3}x\left[\left( \vec{\nabla} \phi_2(\mathbf{x}) \right)^{2}-\left(\vec{\nabla} \phi_1(\mathbf{x}) \right)^{2} \right]\nonumber\\
&&+\frac{1}{2 }  \int_{\mathbb{R}^{3}}  d^{3}x \left[ \left( \vec{\nabla} \hat{\phi}(\mathbf{x}) \right)^{2}+ m_{\eta'}^2  \hat{\phi}^2(\mathbf{x})\right] 
   ~~~~~ \ee
with the  $a'(\mathbf{x})$ field   expressed in terms of the new fields  $\phi_1(\mathbf{x})$ and $\hat{\phi}(\mathbf{x})$ as  
\be
\label{a-ghost}
a'(\mathbf{x})\equiv \frac{1}{m_{\eta'}}\left(\hat{\phi}(\mathbf{x})-\phi_1(\mathbf{x})\right), 
\ee
 while the topological density $q(\mathbf{x})$  operator is expressed in terms of these fields as  
 \be
 \label{q_ghost}
 q =\sqrt{\frac{\zeta}{NL^2}} \vec{\nabla}^2a' =\sqrt{\frac{\zeta}{NL^2m_{\eta'}^2}} \vec{\nabla}^2\left(\hat{\phi}-\phi_1 \right).
 \ee

 This redefinition obviously leads to our previous result (\ref{QCD_final}), (\ref{QCD_ghost}) when we use the Green's functions determined by the Lagrangian (\ref{S_ghost})  for the physical massive field $\hat{\phi}$ and the ghost $\phi_1$, 
  \be
\label{QCD_final_g}
\la q(\mathbf{x}), q(\mathbf{0})\ra_{QCD} =\frac{\zeta}{NL^2}\left[ \delta(\bold{x})-m_{\eta'}^2\frac{e^{-m_{\eta'}r}}{4\pi r} \right]. 
\ee
    
 An important  point here is that the contact term in this framework is explicitly saturated by the topological non-propagating auxiliary fields expressed in terms of the ghost field $\phi_1$, similar to the  Kogut -Susskind (KS) ghost \cite{KS} in two dimensional QED,  or the   Veneziano ghost \cite{ven} in four-dimensional QCD. 
 From our original formulation \cite{Thomas:2011ee} without  any 
auxiliary fields   it is quite obvious that our theory is unitary and causal. When we introduce  the auxiliary fields (which are extremely useful when one attempts  to study the long range order) the unitarity, of course, still holds.    Formally, 
the   unitary holds  in this formulation because  the ghost field $\phi_1$ is always paired up with $\phi_2$ in   every gauge invariant matrix element as explained in ~\cite{KS} (with the only exception being the topological density operator (\ref{q_ghost}) which requires a special treatment presented in  this section).  The condition that enforces this statement is the Gupta-Bleuler-like condition on the physical Hilbert space ${\cal H}_{\mathrm{phys}}$ which reads  
\be\label{gb}
(\phi_2 - \phi_1)^{(+)} \left|{\cal H}_{\mathrm{phys}}\right> = 0 \, ,
\ee
where the $(+)$ stands for the positive frequency Fourier components of the quantized fields.  

$\bullet$The crucial point here is that the formulation of the theory using the topological fields has an enormous advantage as the long range dynamics  is explicitly accounted for in the formulation (\ref{eta-action}) and therefore, in  the equivalent formulation   in terms of the ghost field (\ref{S_ghost}). In solvable ``deformed QCD" model it was a question of taste which framework to choose.   For our studies of superfluidity developed in sections \ref{susceptibility}, \ref{numerics} this is not a   question of taste, but necessity. This is  because an explicit studies  of the vortex network which require  an accounting for   all the configurations with varies geometries and topologies of knotted, wrinkled and  folded   vortices is simply not  technically feasible. At the same time, the studies with the topological auxiliary fields give, at least, qualitative picture of the system. 
The computations in ``deformed QCD" model (where the explicit computations have been    carried out) presented in this Appendix give us some confidence that our formal manipulations with the path integral with topological auxiliary fields in sections \ref{susceptibility}, \ref{numerics} capture  the basic features of the system.


\end{document}